\documentclass[preprint,12pt,authoryear]{elsarticle}
\usepackage[utf8]{inputenc}

\usepackage{amssymb}

\journal{Ocean Modelling}

\usepackage{amsbsy}
\usepackage{amsmath}
\usepackage{amsfonts}
\usepackage{amsthm}
\usepackage{array}
\usepackage{graphicx}
\usepackage{bm}
\usepackage{algorithm}
\usepackage{algorithmic}

\newcommand{\pd}[2]{\frac{\partial #1}{\partial #2}}

\newcommand{\bnabla}{\bm{\nabla}}
\newcommand{\vint}[2][K]{\Big\langle #2  \Big\rangle_{\!\!#1}}

\newcommand{\sint}[2][e]{\Big\langle\!\!\Big\langle #2 \Big\rangle\!\!\Big\rangle_{\!\!#1}}

\newcommand{\mean}[1]{\{\!\!\{#1\}\!\!\}}
\newcommand{\jump}[1]{[\![#1]\!]}

\newcommand{\bu}{\bm{u}}

\newcommand{\bbaru}{\bar{\bm{u}}}

\newcommand{\bphi}{\bm{\phi}}
\newcommand{\Dt}{\Delta t}

\graphicspath{{.}{./figures/}{../figures/}}

\begin{document}

\begin{frontmatter}



\title{Discontinuous Galerkin discretization for two-equation turbulence closure model}


\author{Tuomas K\"arn\"a\corref{cor1}}
\ead{tuomas.karna@fmi.fi}

\address{Finnish Meteorological Institute, Helsinki, Finland}

\cortext[cor1]{Corresponding author}
\address{}

\begin{abstract}
Accurate representation of vertical turbulent fluxes is crucial for numerical ocean modelling, both in global and coastal applications. The state-of-the-art approach is to use two-equation turbulence closure models which introduces two dynamic equations to the system. Solving these equations numerically, however, is challenging due to the strict requirement of positivity of the turbulent quantities (e.g., turbulence kinetic energy and its dissipation rate), and the non-linear source terms that may render the numerical system unstable. In this paper, we present a Discontinuous Galerkin (DG) finite element discretization of the Generic Length Scale (GLS) equations designed to be incorporated in a DG coastal ocean model, Thetis. To ensure numerical stability, the function space for turbulent quantities must be chosen carefully. In this work, we propose to use zeroth degree elements for the turbulent quantities and linear discontinuous elements for the tracers and velocity. The spatial discretization is completed with a positivity preserving semi-implicit time integration scheme. We validate the implementation with standard turbulence closure model benchmarks and an idealized estuary simulation. Finally, we use the full three-dimensional model to simulate the Columbia River plume. The results confirm that the coupled model generates realistic vertical mixing, and remains stable under strongly stratified conditions and strong tidal forcing. River plume characteristics are well captured.
\end{abstract}


\end{frontmatter}


\section{Introduction} \label{sec:intro}

Ocean models rely on parametrizations to account for
sub-grid scale vertical mixing processes.
The relevant eddy coefficients are obtained by means of turbulence closure
models.
While in some applications simple parametrizations, such as algebraic
expressions of eddy viscosity and diffusivity, can be sufficient, in general
more sophisticated schemes are needed in order to model the space-time
evolution of turbulent fluxes.

Representing the dynamic evolution of turbulent fields is crucial especially in
coastal, buoyancy-driven flows.
Vertical mixing plays a major role in bottom boundary layer dynamics, evolution of stratification, and
formation of the surface mixed layer, for example.
It is also essential feature in estuarine circulation and river plume dynamics, which in general cannot be simulated without a sophisticated turbulence closure model.

Zero-equation parametrizations, such as \cite{pacanowski1981}, the K-Profile Parametrization \citep[KPP; ][]{large1994,vanroekel2018}, and the recent ePBL parametrization \citep{reichl2018} are widely used in ocean models due to their simplicity and low computational cost.
These methods do not include a prognostic turbulent variable but parametrize the eddy viscosity as a function of the mean flow state.
Consequently, they are of limited applicability:
the KPP and ePBL models only parametrize the  the surface boundary layer, and none of the schemes can represent the temporal dynamics of turbulence.

The most sophisticated turbulence closures are two-equation models.
These models consist of two partial differential equations, one for the
turbulent kinetic energy (TKE), and another one for an auxiliary turbulent variable
that defines the turbulent length scale.
Such models include the Mellor-Yamada (level 2.5) model \citep{mellor1982},
$k-\varepsilon$ \citep{rodi1987},
$k-\omega$ model \citep{wilcox1988},
and the Generic Length Scale \citep[$gen$;][]{umlauf2003b} model.
The benefit of the GLS formulation is that all of the above closures can
be obtained merely by changing parameters.

The turbulent eddy viscosity and diffusivity are obtained from the state
variables, scaled by so-called (non-dimensional) stability functions.
The most common stability functions are full-equilibrium functions
by \cite{canuto2001} (Canuto A and B), and by \cite{cheng2002}.
Quasi-equilibrium functions, such as by \citep{kantha1994}, are also often used
but have shown to be of limited applicability \citep{umlauf2005}.

Many existing ocean models implement two-equation turbulence closure models.
The generic GLS model has become popular in recent years;
it has been implemented in
SELFE \citep{zhang2008},
SCHISM \citep{zhang2016},
ROMS \citep{warner2005}
and NEMO \citep{reffray2015}
, for example, in
addition to its original implementation in the GOTM library
\citep{burchard1999}.
In most cases, the GLS equations have been implemented in a 1D vertical finite difference or finite volume context (e.g., GOTM);
There have been only a few finite element implementations of two-equations turbulence closure models \citep[e.g.,][]{hill2012}.

In this paper, we present a finite element discretization of the GLS equations
intended to be incorporated in a Discontinuous Galerkin (DG) three-dimensional circulation model.
The $k-\varepsilon$, $k-\omega$, and $gen$ models are tested with a series of standard benchmark test cases, and a realistic simulation of the Columbia River plume.
We consider both the Canuto and Cheng stability functions.
For the sake of simplicity, we do not consider convective adjustment methods.

The weak formulation of the GLS equations in a three-dimensional domain is presented.
The function space for turbulent quantities must be chosen carefully to ensure numerical stability of the coupled turbulence–mean flow system \citep{burchard2002,karna2012}.
We propose using a zeroth order DG space for the turbulent quantities as a natural choice for a linear DG hydrodynamical model;
this choice also allows a straightforward implementation of the nonlinear source terms.
The spatial discretization is completed by a positivity preserving, semi-implicit time integration scheme.
Positivity of the turbulent quantities is ensured by using a $l$-stable implicit solver and Patankar treatment of the source terms.
We present a time integration scheme for the full coupled turbulence–mean flow system.

The GLS model is implemented in the Thetis ocean model \citep{karna2018a}.
Thetis is built on the Firedrake finite element modeling framework \citep{rathgeber2016} which uses a domain-specific language (Unified Form Language; \citealt{alnaes2014}) to
describe the weak forms; A just-in-time code generator is used produce computationally efficient C kernels at runtime.
As such, Firedrake offers high flexibility in terms of describing the mathematical formulation without sacrificing computational efficiency.
It also provides flexible support for various high performance computing platforms, as the code generator can apply hardware specific optimizations automatically.
The present work demonstrates that turbulence closure models can be implemented in such domain specific language frameworks.

The GLS equations are presented
in Section \ref{sec:continuous_eqns}.
The positivity preserving time integration scheme is presented in Section
\ref{sec:time_discretization}, followed by the finite element discretization in
\ref{sec:dg_discretization}.
Test cases are presented in Section \ref{sec:results}, including a
realistic application to the Columbia River plume in
Section\ref{sec:results_creplume}, followed by Discussion and Conclusions in Sections \ref{sec:discussion} and \ref{sec:conclusions}.

\section{Two equation turbulence closure models} \label{sec:continuous_eqns}

\begin{figure}[ht]
  \centering
  \noindent\includegraphics[width=0.5\textwidth]{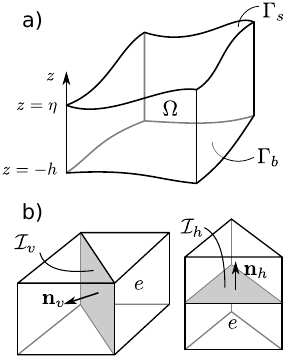}
  \caption
  {
  a) Illustration of the three-dimensional model domain $\Omega$; b) the definition of horizontal and vertical facets, $\mathcal{I}_h$ and $\mathcal{I}_v$, in the extruded three-dimensional mesh.
  }\label{fig:domain}
\end{figure}

\subsection{Generic Length Scale equations} \label{sec:methods_gls}

Let $\Omega$ be the three-dimensional domain in Cartesian coordinates $(x, y, z)$ (Figure \ref{fig:domain} a).
$\Omega$ spans from the sea floor $z=-h(x,y)$ to the free surface $z=\eta(x,y)$;
the bottom and top surfaces are denoted by $\Gamma_b$ and $\Gamma_s$, respectively.

The Generic Length Scale (GLS) turbulence closure model \citep{umlauf2003b} solves the two
equations for turbulent kinetic energy (TKE), $k$, and an auxiliary turbulent variable, $\Psi$,
\begin{align}
  \pd{k}{t} + \bnabla_h \cdot (\bu k) + \pd{(w k)}{z} &= \pd{}{z}\left(\frac{\nu}{\sigma_k} \pd{k}{z}\right) + P + B - \varepsilon, \label{eq:k_eq} \\
  \pd{\Psi}{t} + \bnabla_h \cdot (\bu \Psi) + \pd{(w \Psi)}{z} &= \pd{}{z}\left(\frac{\nu}{\sigma_\Psi} \pd{\Psi}{z}\right) + \frac{\Psi}{k}(c_1 P + c_3 B - c_2 \varepsilon),  \label{eq:psi_eq}
\end{align}
where $t$ denotes the time, $\bu = (u, v, 0)$ and $w$ are the horizontal and vertical velocity, respectively,
and $\bnabla_h = (\partial/\partial x, \partial/\partial y, 0)$ is the horizontal gradient operator.
The vertical eddy viscosity is denoted by $\nu$;
$\sigma_k$ and $\sigma_\Psi$ are the Schmidt numbers for the two state variables, respectively.

The second term in equations \eqref{eq:k_eq}-\eqref{eq:psi_eq} represents horizontal advection.
In many ocean models this term is neglected by assuming horizontal homogeneity of turbulence.
In the coastal ocean, however, turbulence can change drastically in short distances, and this assumption is not valid.
A river plume is a good example:
Turbulent conditions change drastically across the plume front which separates the strongly stratified freshwater plume and weakly stratified coastal waters.
As the front advances in the surface layer, failing to transport the turbulent quantities with the front would inevitably lead into erroneus mixing at the plume front.

The source terms in \eqref{eq:k_eq} are production of TKE, $P$; buoyancy production, $B$; and TKE dissipation rate $\varepsilon$.
$P$ and $B$ depend on the vertical shear frequency $M$ and buoyancy (Brunt–Väisälä) frequency $N$, respectively:
\begin{align}
  P &= \nu M^2, \\
  B &= - \nu' N^2, \\
  M^2 &= \left(\pd{u}{z}\right)^2 + \left(\pd{v}{z}\right)^2, \label{eq:M2_def} \\
  N^2 &= -\frac{g}{\rho_0}\pd{\rho}{z}, \label{eq:N2_def} 
\end{align}
where $\nu'$ is the eddy diffusivity of tracers,
$g$ is the gravitational acceleration, $\rho$ is the density of the water, and $\rho_0$ denotes a constant reference density.

The term $P$ is always positive indicating that vertical velocity shear generates turbulence; $P$ transforms kinetic energy to turbulent kinetic energy \citep{burchard2002}.
The sign of $B$, on the other hand, depends on $N^2$: for stable stratification ($N^2>0$), $B$ is negative and transforms turbulent kinetic energy to potential energy.
Stratification thus inhibits turbulence.
In the case of unstable stratification, however, $B$ is positive and transforms potential energy to turbulence.
\cite{burchard2002} showed that the numerical implementation of $B$ and $P$ should be energy conservative, such that increase of TKE by $P$, for example, matches the corresponding loss of kinetic energy.
The dissipation term, $\varepsilon$, is always positive and represents loss of TKE into heat.

The auxiliary turbulent variable, $\Psi$, is defined by the parameters $p, m, n$:
\begin{align}
  \Psi = \left(c_\mu^0 \right)^p k^m l^n, \label{eq:psi_definition}
\end{align}
where $c_\mu^0$ is a dimensionless empirical parameter, and $l$ is a turbulent length scale.
The source terms of $\Psi$ in \eqref{eq:psi_eq} also depend on $P$, $B$, and $\varepsilon$, but are scaled by $\Psi/k$ and the empirical constants $c_1$, $c_2$, and $c_3$ which depend on the chosen closure.

The parameter $c_3$ controls buoyancy production of $\Psi$.
It is split into two different values for stable and unstable stratification, respectively:

\begin{align}
 c_3 &= \left\{
   \begin{array}{l}
     c_3^{-},\quad \forall N^2 \geq 0 \\
     c_3^{+},\quad \forall N^2 < 0
   \end{array}
 \right.
\end{align}

The parameter $c_3^{+}$ is used to control mixing in unstably stratified conditions;
for the closures considered herein $c_3^{+} = 1.0$.
The value of $c_3^{-}$ depends on the closure as discussed in \ref{sec:choosing_c3minus}.

The TKE dissipation rate, $\varepsilon$, and turbulent length scale, $l$, are computed diagnostically from the state variables:
\begin{align}
  \varepsilon &= \left(c_\mu^0\right)^3 \frac{k^{3/2}}{l} = \left(c_\mu^0\right)^{3 + p/n} k^{3/2 + m/n} \Psi^{-1/n} \label{eq:epsilon_def}, \\
  l &= \left(c_\mu^0\right)^3 \frac{k^{3/2}}{\varepsilon} = \left(c_\mu^0\right)^{-p/n} k^{-m/n} \Psi^{1/n}. \label{eq:len_def}
\end{align}

Based on \eqref{eq:epsilon_def} it is clear that choosing $p=3, m=3/2, n=-1$ yields $\Psi=\varepsilon$ and results in the $k-\varepsilon$ model \citep{rodi1987}.

Finally, the eddy viscosity and diffusivity are given by
\begin{align}
  \nu &= c_\mu k^2/\varepsilon,  \label{eq:nu_def} \\
  \nu' &= c'_{\mu} k^2/\varepsilon, \label{eq:nuprime_def}
\end{align}
respectively, where $c_\mu$ and $c'_\mu$ are non-dimensional stability functions that depend on $N^2$ and $M^2$.
In this work, we consider three different stability functions, Canuto A and B, and Cheng, as defined in \ref{sec:stability-functions}.

The values of the empirical parameter $c_\mu^0$ and the Schmidt number $\sigma_\Psi$ are discussed in \ref{sec:cmu0} and \ref{sec:sigmapsi}, respectively.
The used parameter values are listed in Table \ref{tab:gls_parameters}.

\subsubsection{Boundary conditions for $k$ and $\Psi$}\label{boundary-conditions}

Boundary conditions at the surface and bottom boundaries can be derived from law of the wall conditions.
Within the boundary layer, the turbulent length scale grows linearly with the distance to the boundary, i.e.,

\begin{align}
 l_s &= \kappa ((\eta - z) + z_{0,s}), \label{eq:l_surface} \\
 l_b &= \kappa ((h + z) + z_{0,b}), \label{eq:l_bottom}
\end{align}
where $\kappa=0.4$ denotes the von Karman constant and $z_{0,s}$ and $z_{0,b}$ are the corresponding roughness length scales.
TKE in the (unresolved) boundary layer is \citep{burchard1999}:
\begin{align}
 k_s & = \frac{(u_s^*)^2}{(c_\mu^0)^2} \label{eq:k_surface}\\
 k_b & = \frac{(u_b^*)^2}{(c_\mu^0)^2} \label{eq:k_bottom}
\end{align}
where $u_s^*$ and $u_b^*$ are the friction velocities. From \eqref{eq:k_surface}-\eqref{eq:k_bottom} the Neumann condition for $k$ can be derived,
\begin{align}
  \left. \left( \frac{\nu}{\sigma_k}\pd{k}{z} \right) \right|_{\Gamma_b \cup \Gamma_s} &= 0,
\end{align}
which is used as the boundary condition for $k$.
Note that this is a natural boundary condition in DG finite element formulation as the (diffusive) boundary term vanishes if $\partial k/\partial z = 0$.

Substituting \eqref{eq:l_surface}-\eqref{eq:l_bottom} to \eqref{eq:psi_definition} yields
\begin{align}
 \Psi_s &= \left(c_\mu^0 \right)^p k^m \kappa^n ((\eta - z) + z_{0,s})^n \\
 \Psi_b &= \left(c_\mu^0 \right)^p k^m \kappa^n ((h + z) + z_{0,b})^n
\end{align}

Differentiating with respect to $z$ gives

\begin{align}
  \pd{\Psi_s}{z} &= - n (c_\mu^0)^p k^m \kappa^n ((\eta - z) + z_{0,s})^{n-1} \\
  \pd{\Psi_b}{z} &= n (c_\mu^0)^p k^m \kappa^n ((h + z) + z_{0,b})^{n-1}
\end{align}

In practice, the gradients cannot be evaluated right at the boundary because $\Psi$ can be changing rapidly.
We evaluate the boundary conditions at distance $\Delta z/2$ from the boundary where $\Delta z$ is the vertical element size.
This choice is consistent with most finite volume implementations, and results in stable behavior.
The $\Psi$ boundary conditions then become

\begin{align}
  \left. \left( \frac{\nu}{\sigma_\Psi}\pd{\Psi}{z} \right) \right|_{\Gamma_s} &= - n \frac{\nu}{\sigma_\Psi}(c_\mu^0)^p k^m \kappa^n (\Delta z/2 + z_{0,s})^{n-1} \\
  \left. \left( \frac{\nu}{\sigma_\Psi}\pd{\Psi}{z} \right) \right|_{\Gamma_b} &= n \frac{\nu}{\sigma_\Psi}(c_\mu^0)^p k^m \kappa^n (\Delta z/2 + z_{0,b})^{n-1}.
\end{align}

\subsubsection{Boundary conditions for the momentum equation}\label{sec:mom_boundary-conditions}

Regarding the momentum equation (see \citealt{karna2018a}), we impose the log-layer bottom friction condition on the bottom boundary:

\begin{align}
  \left. \left( \nu \pd{\bu}{z} \right) \right|_{\Gamma_b} &= C_d  \| \bu_b \| \bu_b, \\
 C_d &= \frac{\kappa^2}{\log(\frac{z_b + h + z_0^b}{z_0^b})^2 },
\end{align}

where $C_d$ is the drag coefficient, $z_0$ is the bottom roughness length, $z_b$ is the $z$-coordinate at the middle
of the bottom most element, and $\bu_b = \bu(z_b)$.

On the free surface, wind stress ($\bm{\tau}_w$) is imposed:
\begin{align}
  \left. \left( \nu \pd{\bu}{z} \right) \right|_{\Gamma_s} &= \frac{\bm{\tau}_w}{\rho_0}.
\end{align}
In applications with atmospheric wind forcing, the stress is computed using the formulation by \cite{large1981}, and a constant air density, $\rho_{air} = 1.22\ \text{kg}\ \text{m}^{-3}$.

For tracers, impermeable boundary conditions, with zero advective and diffusive fluxes, are imposed at the surface and bottom boundaries.

\subsection{Steady state solution}

In a steady state, and assuming homogeneous turbulence, the source terms of the governing equations \eqref{eq:k_eq}-\eqref{eq:psi_eq} balance each other resulting in

\begin{align}
 P + B - \varepsilon &= 0 \label{eq:steady_k_source}\\
 c_1 P + c_3 B - c_2 \varepsilon &= 0 \label{eq:steady_psi_source}
\end{align}

It is convenient to define non-dimensional shear and buoyancy frequencies \citep{burchard2001},

\begin{align}
  \alpha_M &= \frac{k^2}{\varepsilon^2} M^2, \\
  \alpha_N &= \frac{k^2}{\varepsilon^2} N^2.
\end{align}

Using the gradient Richardson number,

\begin{align}
  R_i = \frac{N^2}{M^2} =\frac{\alpha_N}{\alpha_M}, \label{eq:gradient_ri}
\end{align}

the equilibrium condition \eqref{eq:steady_k_source}-\eqref{eq:steady_psi_source} can be expressed as \citep{umlauf2003b}

\begin{align}
  R_i^{st} = \frac{c_\mu(R_i^{st})}{c'_\mu(R_i^{st})} \frac{c_2 - c_1}{c_2 - c_3^-}, \label{eq:ri_steady_state}
\end{align}

where $c_\mu$ and $c'_\mu$ are the stability functions, and $R_i^{st}$ denotes the steady state gradient Richardson number (commonly set to value 0.25).

\subsection{Length scale limitation}

\cite{galperin1988} suggested to limit $l$ as

\begin{align}
 l \leq l_{max} = c_{lim} \frac{\sqrt{2k}}{N}.
\end{align}

\cite{umlauf2005} show that the limiting factor $c_{lim}$ is a function of the steady-state Richardson number:
 
\begin{align}
 c_{lim} &= \frac{(c_\mu^0)^{3}}{\sqrt{2}}\sqrt{\alpha_N(R_i^{st})} \label{eq:galperin_clim}
\end{align}

The length scale limitation can be formulated as a limit on $\Psi$ \citep{warner2005}

\begin{align}
 \Psi^{1/n} &\le \sqrt{2} c_{lim} (c_\mu^0)^{p/n} k^{m/n + 1/2} N^{-1} \label{eq:galperin_psi}.
\end{align}

Note that for negative $n$, \eqref{eq:galperin_psi} imposes a lower limit on $\Psi$.
In this work we limit $\Psi$ only, i.e., no limit is applied on $l$ or $\varepsilon$.

\begin{table}[th!]
\caption{
List of parameter values for three different turbulence closure models.
$k-\varepsilon$ and $k-\omega$ values (indicated by $\dagger$) are from Tables 1 and 2 in \cite{umlauf2003b};
The $gen$ model values (indicated by $\ast$) corresponds to the first line in Table 7 in \cite{umlauf2003b}.
In all cases, $c_{lim}$ is computed with \eqref{eq:galperin_clim};
$c_3^-$, $c_\mu^0$ and $\sigma_\Psi$ are computed as detailed in \ref{sec:choosing_parameters}.
The shown values correspond to the Canuto A stability functions.
}\label{tab:gls_parameters}
\begin{center}
\small
\begin{tabular}{|c|r|r|r|}
\hline
Parameter            & $k-\epsilon$   & $k-\omega$ & $gen$           \\
\hline
$p$                  & 3$^\dagger$    & -1.0$^\dagger$ &  2.0$^\ast$   \\
$m$                  & 1.5$^\dagger$  &  0.5$^\dagger$ &  1.0$^\ast$   \\
$n$                  & -1.0$^\dagger$ & -1.0$^\dagger$ & -0.67$^\ast$  \\
\hline
$\sigma_k$           & 1.0$^\dagger$ & 2.0$^\dagger$   & 0.8$^\ast$    \\
$\sigma_\Psi$        & 1.20$^\dagger$ & 2.072$^\dagger$   & 1.18$^\ast$   \\
\hline
$c_1$                & 1.44$^\dagger$ & 0.555$^\dagger$  & 1.0$^\ast$    \\
$c_2$                & 1.92$^\dagger$ & 0.833$^\dagger$  & 1.22$^\ast$   \\
$c_3^+$              & 1.0            & 1.0              & 1.0           \\
$c_3^-$              & -0.629         & -0.643           & 0.052         \\
\hline
$R_i^{st}$           & 0.25           & 0.25       & 0.25          \\
$c_\mu^0$            & 0.5270         & 0.5270     & 0.5270        \\
$\kappa$             & 0.4            & 0.4        & 0.4           \\
\hline
$c_{lim}$              & 0.267          & 0.267      & 0.267         \\
\hline
$k_{\text{min}}$     & 1.0$\times 10^{-6}$ & 7.6$\times 10^{-6}$ & 1.0$\times 10^{-6}$ \\
$\Psi_{\text{min}}$  & 1.0$\times 10^{-14}$ & 1.0$\times 10^{-14}$ & 1.0$\times 10^{-14}$ \\
\hline
\end{tabular}
\end{center}
\end{table}

\section{Temporal discretization} \label{sec:time_discretization}

This section outlines the temporal discretization before introducing the DG discretization in Section \ref{sec:dg_discretization}.

\subsection{Thetis coastal ocean model} \label{sec:methods_thetis}

The turbulence closure model was implemented in the Thetis three-dimensional
circulation model \citep{karna2018a}.
Thetis solves the hydrostatic equations with a semi-implicit DG finite element
method.
The equations are solved in a time-dependent 3D mesh, unstructured in the
horizontal direction.
The mesh moves in the vertical direction to track the free surface undulation;
the mesh movement is implemented with the Arbitrary Lagrangian–Eulerian (ALE) method.
A second-order split-implicit time integration scheme is used to advance the
equations in time.
Advection of 3D fields is solved with an explicit Strong Stability-Preserving (SSP) scheme.
Vertical diffusion is treated implicitly.
The model formulation is described in detail in \cite{karna2018a}.

For 3D advection, Thetis uses a two-stage, second-order SSP Runge-Kutta scheme, SSPRK(2,2) \citep{shu1988}.
The SSP property ensures that the solution is non-oscillatory as long as the CFL condition is satisfied.
With DG spatial discretization, this implies that the element mean value is positive definite; overshoots can still arise within an element.
To this end, a vertex-based slope limiter \citep{kuzmin2010} is used to redistribute mass in the element if local overshoots are present.
The slope limiter is applied as a post-processing step after solving the equations.
\cite{karna2018a} demonstrate that the model is mass conservative (typical error being $\mathcal{O}(10^{-12})$) and non-oscillatory (largest overshoots are of order $\mathcal{O}(10^{-5})$ in extreme cases).

In the GLS implementation, the
turbulent quantities are advanced in time similarly to other 3D fields:
Horizontal and vertical advection is solved with the
same SSPRK ALE scheme.
Vertical dynamics of $k$ and $\Psi$ are solved implicitly.

Thetis source code is freely available.
We have also archived the exact source code version used to to produce the results in this paper (see \ref{sec:source_code}).

\subsection{Implicit and explicit terms} \label{sec:implicit_explicit_split}

The governing equations \eqref{eq:k_eq}-\eqref{eq:psi_eq} are advanced in time with a fractional step method.
We first separate the explicit advection terms, the remaining implicit terms:
\begin{align}
 \pd{k}{t} + A_k(k, \bu, w) &= D_k(k, \nu) + S_k(k, \varepsilon, \nu, \nu', M, N), \\
 \pd{\Psi}{t} + A_\Psi(\Psi, \bu, w) &= D_\Psi(\Psi, \nu) + S_\Psi(\Psi, k, \varepsilon, \nu, \nu', M, N),
\end{align}
where $A_i,\ i=\{k,\Psi\},$ denote the horizontal and vertical advection terms,
$D_i$ the vertical diffusion terms, and $S_i$ the source terms:
\begin{align}
 S_k &= P + B - \varepsilon \\
 S_\Psi &= \frac{\Psi}{k}(c_1 P + c_3 B - c_2 \varepsilon).
\end{align}

Let $\Dt$ denote the time step.
Given the state variables at time level $n$, the time update of $k$ can be divided into an explicit and implicit steps ($\Psi$ is treated analogously):
\begin{align}
  k^{*} &= k^{n} - \Dt A_k(k^{n}, \bu^{n}, w^{n}), \label{eq:k_adv_step} \\
  k^{n+1} &= k^{*} + \Dt D_k(k^{n+1}, \nu^{n}) + \Dt S_k(k^{n+1}, \varepsilon^{n}, \nu^{n}, \nu'^{n}, M^{n}, N^{n}). \label{eq:k_impl_step}
\end{align}

The explicit advection update \eqref{eq:k_adv_step} is solved with the ALE SSPRK method as described in \cite{karna2018a}.
The implicit update \eqref{eq:k_impl_step} is presented below.
Note that $D_k$ and $S_k$ have been linearized with respect to the prognostic variable: only $k$ is taken at time level $n+1$.

To ensure mass conservation and positivity preservation of the whole scheme, both stages \eqref{eq:k_adv_step} and \eqref{eq:k_impl_step} must satisfy these properties.
As stated above, the 3D advection step is conservative and positivity preserving.
The implicit scheme, however, must be designed with care:
The diffusion operator may introduce oscillations if viscosity (or time step) is very large.
Moreover, the non-linear source terms ($S_k$ and $S_\Psi$) may cause negative values unless special treatment is applied.

\subsection{Implicit solver for the GLS equations} \label{sec:implicit_gls_scheme}

The prognostic variables $k$ and $\Psi$ are positive by definition.
The discretized system must maintain positivity of these variables at all times to ensure physically sound and numerically stable solution.

As the vertical problem is generally stiff, it is crucial to use an L-stable
implicit method to avoid spurious oscillations \citep{alexander1977,hairer1996}.
L-stability means that the amplification factor of the scheme tends to zero as the time step tends to infinity, effectively dampening unresolved high-frequency oscillations.
It is worth noting that the commonly-used Crank-Nicolson method is not L-stable and may lead to erratic and unphysical solutions.
The simplest L-stable choice is the Backward Euler method \eqref{eq:k_impl_step}. It is used in \cite{karna2018a} and also employed in this work.
Alternatively, a more accurate multi-stage, L-stable Diagonally Implicit Runge Kutta scheme could be used \citep[e.g.,][]{ascher1997}.
Practical evaluations, however, showed that using a two-stage second order method did not affect the results significantly but did increase the cost of the implicit solve by a factor of two.

In Thetis, the domain is decomposed into parallel sub-regions in the horizontal direction, i.e., each vertical column of elements lies within a single process.
Because the implicit solve only contains vertical fluxes, the implicit solve is local and does not involve parallel communication.

\subsection{Patankar treatment of source terms} \label{sec:patankar_treatment}

The non-linear source terms must be treated carefully because they can trigger instabilities and/or generate negative values.
Positivity preservation is based on the Patankar treatment where all sink terms (i.e., $S_i < 0$) are treated implicitly \citep{patankar1980,burchard2003}
 to ensure that they cannot overshoot to produce negative quantities during temporal integration.
A similar approach is used in GOTM, for example.

In the $k$ equation, the shear production term, $P$, and TKE dissipation rate, $\varepsilon$, are always positive, resulting in a source and sink term, respectively.
The sign of the buoyancy production term, $B$, on the other hand, depends on the sign of $N^2$.
We therefore split the term in two parts: $B = B_{+} + B_{-}$ with $B_{+} \geq 0$ and $B_{-} \leq 0$:
\begin{align}
 B_{+} &= - \nu' \min(N^2, 0), \\
 B_{-} &= - \nu' \max(N^2, 0).
\end{align}

In the $\Psi$ equation, we have $c_1 P \geq 0$ and $c_2 \varepsilon \geq 0$.
In this case the sign of the $B$ term, however, also depends on the parameter $c_3$: $c_3^+>0$ is used when $B\geq0$, and $c_3^-$ when $B<0$ ($c_3^-$ can be negative).
Denoting $c_3 B = c_3^+ B_+ + c_3^- B_-$, we define:
\begin{align}
 (c_3 B)_{+} &= \max(c_3 B, 0), \\
 (c_3 B)_{-} &= \min(c_3 B, 0).
\end{align}

We can now split the source terms in \eqref{eq:k_impl_step} to explicit and implicit parts,
\begin{align}
  S_k^{n+1} &= (S_k^{+})^n + (S_k^{-})^{n+1}, \\
  (S_k^{+})^{n} &= P^{n} + B_{+}^{n},\\
  (S_k^{-})^{n+1} &= \frac{k^{n+1}}{k^{*}}(B_{-}^{n} - \varepsilon^{n})
\end{align}
Note that in $S_k^{-}$ we have employed the Patankar treatment, i.e., scaled the sink term with the ratio $k^{n+1}/k^{*}$ to render it
implicit (here $k^{*}$ is the ``old'' value due to the fractional stepping).

The update for $\Psi$ is analogous with the source terms:
\begin{align}
  (S_\Psi^{+})^n &= \frac{\Psi^{n}}{k^{n}}(c_1 P^{n} + (c_3 B)^{n}_{+}), \\
  (S_\Psi^{-})^{n+1} &= \frac{\Psi^{n+1}}{k^{n}}((c_3 B)_{-}^{n} - c_2 \varepsilon^{n}).
\end{align}

\subsection{Setting minimum values for $k$ and $\varepsilon$} \label{sec:methods_minvalue}

The presented solver ensures that $k$ and $\Psi$ remain positive throughout the simulation.
However, due to round-off errors small negative values can appear which lead into undefined values (e.g., due to the square root in \eqref{eq:len_def}).
The turbulent quantities are, therefore, cropped to a small positive minimum value; this is a common practice in several implementations
\citep[e.g.,][]{umlauf2004,warner2005}.
The minimum values for $k$ and $\Psi$ are listed in Table \ref{tab:gls_parameters}.
In addition we impose minimum values $l_{min}=1.0\times 10^{-12}\ \text{m}$ and $\nu_{min} = \nu'_{min} = 1.0\times 10^{-8}\ \text{m}^2\ \text{s}^{-1}$.
It should be noted that the user can modify these limits and also introduce additional background viscosity/diffusivity if needed.
These limits are imposed after every update.

\section{Spatial discretization} \label{sec:dg_discretization}

\subsection{Mathematical notation} \label{sec:dg_notation}

The domain $\Omega$ is divided into three-dimensional elements $e \in \mathcal{P}$.
The mesh is generated by extruding a two-dimensional surface mesh over the vertical dimension.
The surface mesh consist of either triangles or quads, resulting in triangular prisms, or hexahedral 3D elements, respectively.

We denote discontinuous Galerkin function spaces of degree $p$ by $\text{P}^{\text{DG}}_p$.
A function $f \in \text{P}^{\text{DG}}_p$ is a polynomial of degree $p$ in each element and discontinuous at the element interfaces.
On the extruded mesh, we define function spaces as a Cartesian product of the horizontal (2D) and vertical (1D) elements: $\text{P}^{\text{DG}}_p\times\text{P}^{\text{DG}}_q$;
a function in this space is therefore a polynomial of degree $p$ in the horizontal direction and degree $q$ in the vertical direction.

The set of horizontal and vertical interfaces are denoted by $\mathcal{I}_h$ and $\mathcal{I}_v$, respectively (Figure \ref{fig:domain} b).
The outward unit normal vector is denoted by $\mathbf{n}= (n_x, n_y, n_z)$; its restriction on the horizontal and vertical interfaces are denoted by $\mathbf{n}_h$, and $\mathbf{n}_v$, respectively.
The vertical facets $\mathcal{I}_v$ are strictly vertical, implying $\mathbf{n}_v = (n_x, n_y, 0)$.

In the weak forms we use the following notation for volume and interface integrals

\begin{align*}
 \vint[\Omega]{\bullet} &= \int_{\Omega} \bullet\; \mathrm{d}\mathbf{x}, \\
 \sint[\partial\Omega]{\bullet} &= \int_{\partial\Omega} \bullet\; \mathrm{d}s.
\end{align*}

In interface terms we additionally use the average $\mean{\cdot}$ and jump $\jump{\cdot}$ operators for arbitrary scalar $a$ and vector $\mathbf{u}$ fields:

\begin{align*}
 \mean{a} &= \frac{1}{2}(a^+ + a^-), \\
 \mean{\bu} &= \frac{1}{2}(\mathbf{u}^+ + \mathbf{u}^-), \\
 \jump{a \mathbf{n}} &= a^+ \mathbf{n}^+ + a^- \mathbf{n}^-, \\
 \jump{\mathbf{u} \cdot \mathbf{n}} &= \mathbf{u}^+ \cdot \mathbf{n}^+ + \mathbf{u}^- \cdot \mathbf{n}^-,
\end{align*}
where the superscripts '$+$' and '$-$' arbitrarily label the values on
either side of the interface.
Note that for the outward normal vectors it holds $\mathbf{n}^+ = -\mathbf{n}^-$.

\subsection{Choosing function spaces} \label{sec:dg_function_spaces}

The prognostic variables of the GLS system \eqref{eq:k_eq}-\eqref{eq:psi_eq} are $k$ and $\Psi$.
Diagnostic variables include the dissipation rate $\varepsilon$, turbulent length scale $l$, vertical eddy viscosity $\nu$, and diffusivity $\nu'$.
Diagnostic variables originating from the 3D circulation model are the vertical shear frequency $M$ and buoyancy frequency $N$.
The choice of the function spaces where these variables reside is crucial for numerical stability and accuracy.

The production terms $P$ and $B$ are of crucial importance as they can excite spurious modes in the numerical solution.
Because $P = \nu M^2$, the function spaces of $\nu$ and $M$ determine the polynomial degree of $P$:
for example, if $\nu$ and $M$ are linear, $P$ is nominally cubic (but could be projected to a lower function space).

For numerical stability it is natural to choose function spaces that can satisfy the steady state condition $P + B - \varepsilon = 0$ point-wise (i.e., in a strong sense instead of an integral, weak sense).
This implies that $P$, $B$, and $\varepsilon$ should belong to the same function space. As we can have $\Psi=\varepsilon$ it is convenient to choose that $k$ and $\Psi$ also belong to the same space.
This is also desirable because it ensures that the source terms can be directly mapped to the space of the prognostic variables, and cannot thus excite spurious modes.

Thetis uses linear discontinuous Galerkin elements for 3D tracers and all velocity components.
That is, the variables belong to a function space, $\mathcal{W}_1 = \text{P}^{\text{DG}}_1\times\text{P}^{\text{DG}}_1$.
Consequently, $\partial u/\partial z$ belongs to $\text{P}^{\text{DG}}_0$ space in the vertical direction.
Noting that a product of two functions in $\text{P}^{\text{DG}}_0$ still belongs to the same space,
$\text{P}^{\text{DG}}_0$ is a natural choice for the shear frequency $M$ as well.
The same reasoning holds also for the tracer gradients and squared buoyancy frequency $N^2$ if one assumes a linear equation of state.

The above requirements can be satisfied by choosing that the turbulent quantities belong to the degree zero DG space, $\mathcal{W}_0 = \text{P}^{\text{DG}}_0\times\text{P}^{\text{DG}}_0$.
With this choice we in fact have $f, g \in \mathcal{W}_0 \Rightarrow fg, f/g, f\sqrt{g} \in \mathcal{W}_0$ which implies that the stability functions and diagnostic variables can be computed point-wise.
The function spaces are summarized in Table \ref{tab:functionspaces}.

It would be desirable to use the same $\mathcal{W}_1$ space for $k$ and $\Psi$ as well, but (at least in the case of $k-\varepsilon$ model) this choice violates the $P = \varepsilon$ steady state condition and leads to unstable solutions.
Finding higher order function spaces for the turbulent quantities remains an open question.

\subsection{Discontinuous Galerkin discretization of the GLS equations} \label{sec:methods_gls_dg}

\begin{table}[ht!]
\caption{
Prognostic and diagnostic variables and their function spaces.
The shear and buoyancy frequencies, $M$ and $N$ are computed diagnostically from the state variables of the hydrodynamical model, and fed to the GLS model as inputs.
Viscosity and diffusivity, $\nu$ and $\nu'$, on the other hand, are fed to the hydrodynamical model.
}\label{tab:functionspaces}
\begin{center}
\begin{tabular}{cccc}
\multicolumn{4}{l}{\textbf{GLS Model}} \\ 
\multicolumn{4}{c}{Prognostic variables} \\ \hline
Field & Symbol & Equation & Function space \\ \hline
Turbulent kinetic energy & $k$ & \eqref{eq:k_eq} & $\text{P}^{\text{DG}}_0\times\text{P}^{\text{DG}}_0$ \\
Aux. turbulent variable & $\Psi$ & \eqref{eq:psi_eq} & $\text{P}^{\text{DG}}_0\times\text{P}^{\text{DG}}_0$ \\
\multicolumn{4}{c}{Diagnostic variables} \\ \hline
TKE dissipation rate & $\varepsilon$ & \eqref{eq:epsilon_def} & $\text{P}^{\text{DG}}_0\times\text{P}^{\text{DG}}_0$ \\
Turbulent length scale & $l$ & \eqref{eq:len_def} & $\text{P}^{\text{DG}}_0\times\text{P}^{\text{DG}}_0$ \\
Eddy viscosity & $\nu$ & \eqref{eq:nu_def} & $\text{P}^{\text{DG}}_0\times\text{P}^{\text{DG}}_0$ \\
Eddy diffusivity & $\nu'$ & \eqref{eq:nuprime_def} & $\text{P}^{\text{DG}}_0\times\text{P}^{\text{DG}}_0$ \\
\multicolumn{4}{l}{\textbf{Hydrodynamical model}} \\
\multicolumn{4}{c}{Prognostic variables} \\ \hline
Horizontal velocity & $\boldsymbol{u}$ & (24-25) in TK2018 & $[\text{P}^{\text{DG}}_1\times\text{P}^{\text{DG}}_1]^2$ \\
Temperature & $T$ & (26) in TK2018 & $\text{P}^{\text{DG}}_1\times\text{P}^{\text{DG}}_1$ \\
Salinity & $S$ & (26) in TK2018 & $\text{P}^{\text{DG}}_1\times\text{P}^{\text{DG}}_1$ \\
\multicolumn{4}{c}{Diagnostic variables} \\ \hline
Density & $\rho$ & (14) in TK2018 & $\text{P}^{\text{DG}}_1\times\text{P}^{\text{DG}}_1$ \\
Shear frequency & $M$ & \eqref{eq:M2_def} & $\text{P}^{\text{DG}}_0\times\text{P}^{\text{DG}}_0$ \\
Buoyancy frequency & $N$ & \eqref{eq:N2_def} & $\text{P}^{\text{DG}}_0\times\text{P}^{\text{DG}}_0$
\end{tabular}
\end{center}
\end{table}

Let $\phi \in \mathcal{W}$ be a test function.
The weak formulations are derived by taking the continuous equations (\ref{eq:k_eq}-\ref{eq:psi_eq}), multiplying them by $\phi$, and integrating over the domain $\Omega$.
The weak formulation then reads: find $k,\Psi \in \mathcal{W}$ such that

\begin{align}
 \pd{}{t} \vint[\Omega]{ k \phi } + \mathcal{A}_{h,k} + \mathcal{A}_{v,k} &= \mathcal{D}_k + \mathcal{S}_k ,\ \forall \phi \in \mathcal{W}, \label{eq:weak_k} \\
 \pd{}{t} \vint[\Omega]{ \Psi \phi } + \mathcal{A}_{h,\Psi} + \mathcal{A}_{v,\Psi} &= \mathcal{D}_\Psi + \mathcal{S}_\Psi ,\ \forall \phi \in \mathcal{W}, \label{eq:weak_psi}
\end{align}
where $\mathcal{A}_{h,i}$ and $\mathcal{A}_{v,i}$, $i=k,\Psi$ denote the
horizontal and vertical advection terms,
while $\mathcal{D}_i$ and $\mathcal{S}_i$ are the diffusion and source terms,
respectively.

The advection terms are discretized with a standard first order DG upwind technique
as implemented in the hydrodynamical model \citep{karna2018a}.
The diffusion terms are treated with Symmetric Interior Penalty Galerkin
method \citep[SIPG;][]{epshteyn2007} with a penalty factor $\sigma = 1/L$ where $L$ is
the vertical element size \citep{karna2018a}.

The source terms are split to sources and sinks as shown in Section \ref{sec:patankar_treatment}:
\begin{align}
\mathcal{S}_k
     &= \underbrace{\vint[\Omega]{S_k^+\phi}}_{\mathcal{S}_k^{+}}
     + \underbrace{\vint[\Omega]{ S_k^- \phi}}_{\mathcal{S}_k^{-}}\\
\mathcal{S}_\Psi
     &= \underbrace{\vint[\Omega]{S_\Psi^+ \phi}}_{\mathcal{S}_\Psi^{+}}
     + \underbrace{\vint[\Omega]{S_\Psi^- \phi}}_{\mathcal{S}_\Psi^{-}}
\end{align}

Let $\phi_i$ be the basis of the function space $\mathcal{W}$.
We require that the solution belongs to this space, i.e., $k,\Psi \in \mathcal{W}$, and use the finite element approximations (e.g., $k = \sum_i k_i \phi_i$).
Using the basis functions as test functions, we can write
(\ref{eq:weak_k}-\ref{eq:weak_psi}) in a bilinear form:

\begin{align}
 \pd{}{t}(\mathbf{M} k) + \mathbf{A}_{h,k} + \mathbf{A}_{v,k} &= \mathbf{D}_k + \mathbf{S}_{+,k} + \mathbf{S}_{-,k} \label{eq:matrixform_k},\quad \forall \phi_j \in \mathcal{W}\\
 \pd{}{t}(\mathbf{M} \Psi) + \mathbf{A}_{h,\Psi} + \mathbf{A}_{v,\Psi} &= \mathbf{D}_\Psi + \mathbf{S}_{+,\Psi} + \mathbf{S}_{-,\Psi},\quad \forall \phi_j \in \mathcal{W} \label{matrixform_psi}
\end{align}
where $[\mathbf{M}]_{i,j} = \vint[\Omega]{\phi_i \phi_j}$ is the mass matrix, and the other terms are also bilinear, e.g, $\mathbf{D}_k = \mathbf{D}_k(\phi_i, \phi_j)$.

\subsection{Computing shear and buoyancy frequencies} \label{sec:shear_buoy_freq}

Computing the vertical shear and buoyancy frequencies accurately is crucial for
maintaining numerical stability \citep{karna2012}.
We compute the vertical gradients weakly.
Denoting the shear frequency vector by $\bm{M} = \pd{\bu}{z}$ we define

\begin{align}
  \vint[\Omega]{\bm{M} \cdot \bphi} &= - \vint[\Omega]{\bu \cdot \pd{\bphi}{z}}
  + \sint[\mathcal{I}_h]{\mean{\bu} \cdot \jump{\bphi n_z}}
  + \sint[\Gamma_s \cup \Gamma_b]{\bu \cdot \bphi n_z} \\
  M^2 &= |\bm{M}|^2,
\end{align}

Analogously, we can compute the $\pd{\rho'}{z} = R_z$
\begin{align}
  \vint[\Omega]{R_z \phi} &= -\vint[\Omega]{\rho' \pd{\phi}{z}}
  + \sint[\mathcal{I}_h]{\mean{\rho'} \jump{\phi n_z}}
  + \sint[\Gamma_s \cup \Gamma_b]{\rho' \phi n_z} \\
  N^2 &= -\frac{g}{\rho_0} R_z.
\end{align}

\subsection{Time integration} \label{sec:methods_timeintegration}

For convenience we re-write the $k$ and $\Psi$ weak forms as

\begin{align}
 \pd{k}{t} &= F_k(k, \bu, w) + G_k(k, \varepsilon, \nu, \nu', M, N) \\
 \pd{\Psi}{t} &= F_\Psi(\Psi, \bu, w) + G_\Psi(\Psi, k, \varepsilon, \nu, \nu', M, N)
\end{align}
where $F_k$ and $F_\Psi$ contain the advection terms whilst $G_k$ and $G_\Psi$ contain the vertical diffusion and source terms.

The solution can then be split in explicit advection stage and implicit vertical dynamics stage.
The explicit stage is solved with SSPRK(2,2) scheme \citep{karna2018a}:

\begin{align}
 k^{(1)} &= k^{n} + \Delta t \tilde{F}_k(k^{n}, \bu^{n}, w^{n} - w_m^{(1)}), \\
 \tilde{k}^{n+1} &= k^{n} + \Delta t \frac{1}{2}\left( \tilde{F}_k(k^{n}, \bu^{n}, w^{n} - w_m^{(1)}) + \tilde{F}_k(k^{(1)}, \bu^{(1)}, w^{(1)} - w_m^{n+1}) \right),
\end{align}
where $w_m$ denotes the vertical mesh velocity and $\tilde{F}$ denotes the ALE weak forms of $F_k$ terms (see \cite{karna2018a}).
The advection stage for $\Psi$ is derived analogously.

Using a linearized Backward Euler update, the implicit stage reads:
\begin{align}
 k^{n+1} &= \tilde{k}^{n+1} + \Delta t \tilde{G}_k(k^{n+1}, \varepsilon^{n}, \nu^{n}, \nu'^{n}, M^{n}, N^{n}), \\
 \Psi^{n+1} &= \tilde{\Psi}^{n+1} + \Delta t \tilde{G}_\Psi(\Psi^{n+1}, k^{n}, \varepsilon^{n}, \nu^{n}, \nu'^{n}, M^{n}, N^{n}).
\end{align}
The full time integration scheme is summarized in Algorithm 1.

\begin{algorithm}[ht!]
 \caption{Summary of the coupled time integration algorithm.}\label{alg:time_integration}
 \begin{algorithmic}[1]
    \REQUIRE Model state variables at time $t_n$: $\eta,\bar{\mathbf{u}},T,S,\mathbf{u}', k, \Psi$
    \\ \hspace*{-18pt} \textbf{First stage:} \\
    \STATE Solve 2D system: $\eta^{(1)},\bbaru^{(1)}$
    \STATE Solve explicit 3D ALE problem: $T^{(1)},S^{(1)},\bu'^{(1)},k^{(1)},\Psi^{(1)}$
    \STATE Update 2D coupling term, density, and vertical velocity
    \\ \hspace*{-18pt} \textbf{Second stage:}\\
    \STATE Solve 2D system: $\eta^{n+1},\bbaru^{n+1}$
    \STATE Solve explicit 3D ALE problem: $\tilde{T}^{n+1},\tilde{S}^{n+1},\tilde{\bu}'^{n+1},\tilde{k}^{n+1},\tilde{\Psi}^{n+1}$
    \STATE Update 2D coupling term
    \\ \hspace*{-18pt} \textbf{Final stage:}\\
    \STATE Solve mean flow vertical dynamics implicitly: $T^{n+1}, S^{n+1}, \bu'^{n+1}$
    \STATE Update density, vertical velocity, bottom friction
    \STATE Solve turbulence vertical dynamics implicitly: $\Psi^{n+1}, k^{n+1}$
    \STATE Update diagnostic turbulence fields: $l^{n+1}, \varepsilon^{n+1}, \nu^{n+1}, \nu'^{n+1}$
 \end{algorithmic}
\end{algorithm}

\section{Results} \label{sec:results}

We first verify the GLS implementation with a series of standard turbulence closure
test cases,
followed by a realistic simulation of the Columbia River plume.
All the tests have been carried out with the $k-\varepsilon$, $k-\omega$, and
the $gen$ models as described in Table \ref{tab:gls_parameters}.

\subsection{Bottom friction} \label{sec:results_botfriction}

The first test is an open channel with bottom friction.
The fluid is initially at rest, forced only by a constant free surface slope
$\partial \eta/ \partial x = -1\times10^{-5}$.
In the absence of rotation, the flow converges to a steady state solution where
the pressure gradient is balanced by the bottom friction.

Near the bed the flow velocity
follows the well-known logarithmic profile, which can be expressed as
\citep[e.g.,][]{hanert2007}

\begin{align}
 \bu(z) &= \frac{\bu^*_b}{\kappa} \log \left(\frac{z_0^b + z + h}{z_0^b}\right)
\end{align}

where $\bu^*_b$ is the bottom friction velocity and
$z_0^b = 1.5 \times 10^{-3}\ \text{m}$ is the bottom roughness length.
The bottom friction velocity can be solved from the momentum balance:

\begin{align}
 u^*_b &= \sqrt{gH \left|\pd{\eta}{x}\right|}.
\end{align}

The eddy viscosity follows a parabolic profile

\begin{align}
 \nu &= - \frac{u^*_b}{H} \kappa (z_0^b + z + h) z.
\end{align}

The free flow was simulated in a 5.0 km wide square domain with
constant 15 m depth.
Horizontal mesh resolution was 2.5 km.
Periodic boundary conditions were used in the $x$ direction.
Initially the flow is at rest.
A steady state solution is reached after roughly 12 h of simulation.
All simulations were carried out with 25 s time step.

\begin{figure}[ht]
  \centering
  \noindent\includegraphics[width=1.0\textwidth]{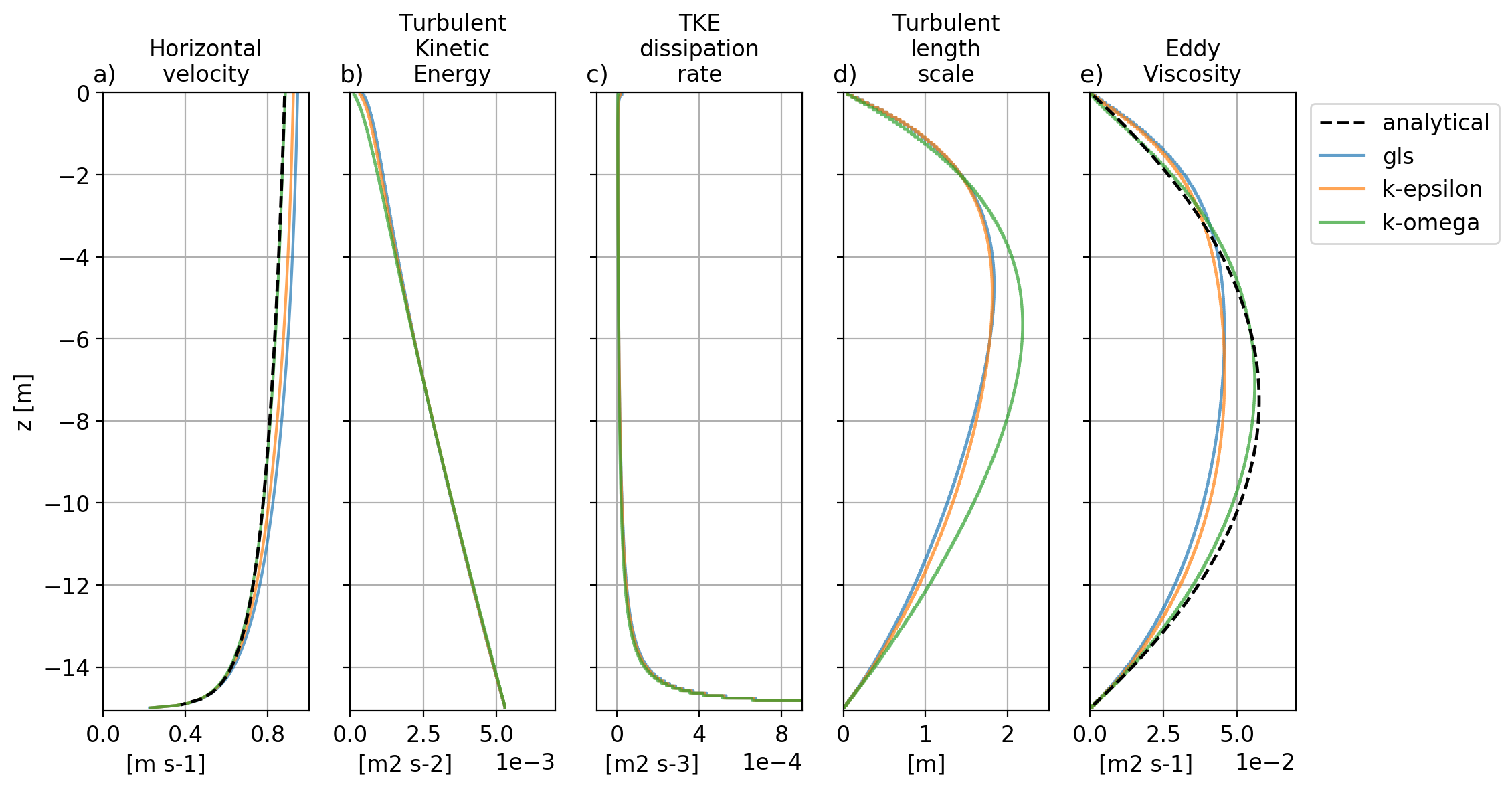}
  \caption
  {Bottom friction test with different turbulence closure models.
  In all cases the Canuto A stability functions and 250 vertical levels were used.
  Black dashed line stands for the analytical velocity and viscosity profiles.
  }\label{fig:bfriction_closure}
\end{figure}

First, we experimented the three closure models,
GLS, $k-\varepsilon$, and $k-\omega$ with Canuto A stability functions.
Vertical profiles of velocity, TKE, $\varepsilon$, $l$, and viscosity are shown
in Figure
\ref{fig:bfriction_closure}.
In all simulations 250 vertical levels were used, resulting in
6 cm vertical resolution.
The parabolic Courant number, $\nu \Delta t/(\Delta z)^2$, was approximately 300.
The velocity profile is close to the analytical logarithmic solution in all
cases.
TKE shows the expected nearly linear profile.
In terms of eddy viscosity, the $k-\omega$ model matches best to the analytical
solution.
GLS and $k-\varepsilon$ models tend to overestimate viscosity in the upper
water column, while underestimating it in the middle.
This behavior is in-line with other results in the literature
\citep{warner2005}.
One possible cause for this deviation is the surface boundary condition
for $\Psi$, which affects both $\Psi$ and $l$ near the free surface (Figure \ref{fig:bfriction_closure} d).

\begin{figure}[ht]
  \centering
  \noindent\includegraphics[width=0.8\textwidth]{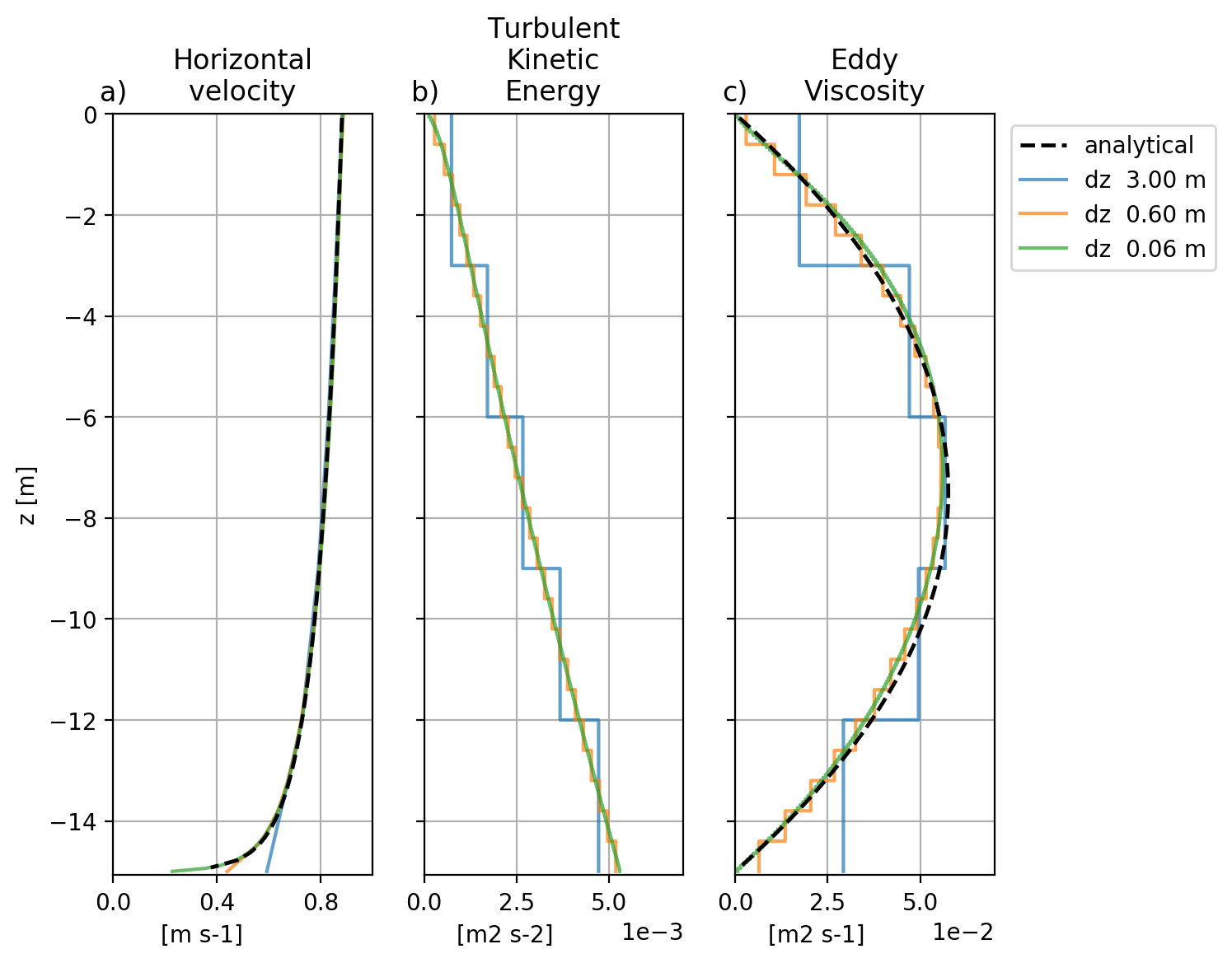}
  \caption
  {Bottom friction test with varying number of levels.
  The $k-\omega$ closure with Canuto A stability functions was used.
  Black dashed line stands for the analytical velocity and viscosity profiles.
  }\label{fig:bfriction_reso}
\end{figure}

Second, the $k-\omega$ model was run with different
vertical grids consisting of 5, 25, and 250 levels (Figure
\ref{fig:bfriction_reso}).
The corresponding parabolic Courant numbers are 0.1, 3, and 300.
The velocity profiles are close to the analytical solution for all the
resolutions.
As resolution is increased viscosity converges to a parabola close to the
analytical solution.
This test demonstrates that the effective friction felt by the water column
does not depend strongly on the vertical resolution.
In addition, it shows that the numerical solver remains stable even with
small vertical elements (6 cm).

\begin{figure}[ht]
  \centering
  \noindent\includegraphics[width=0.8\textwidth]{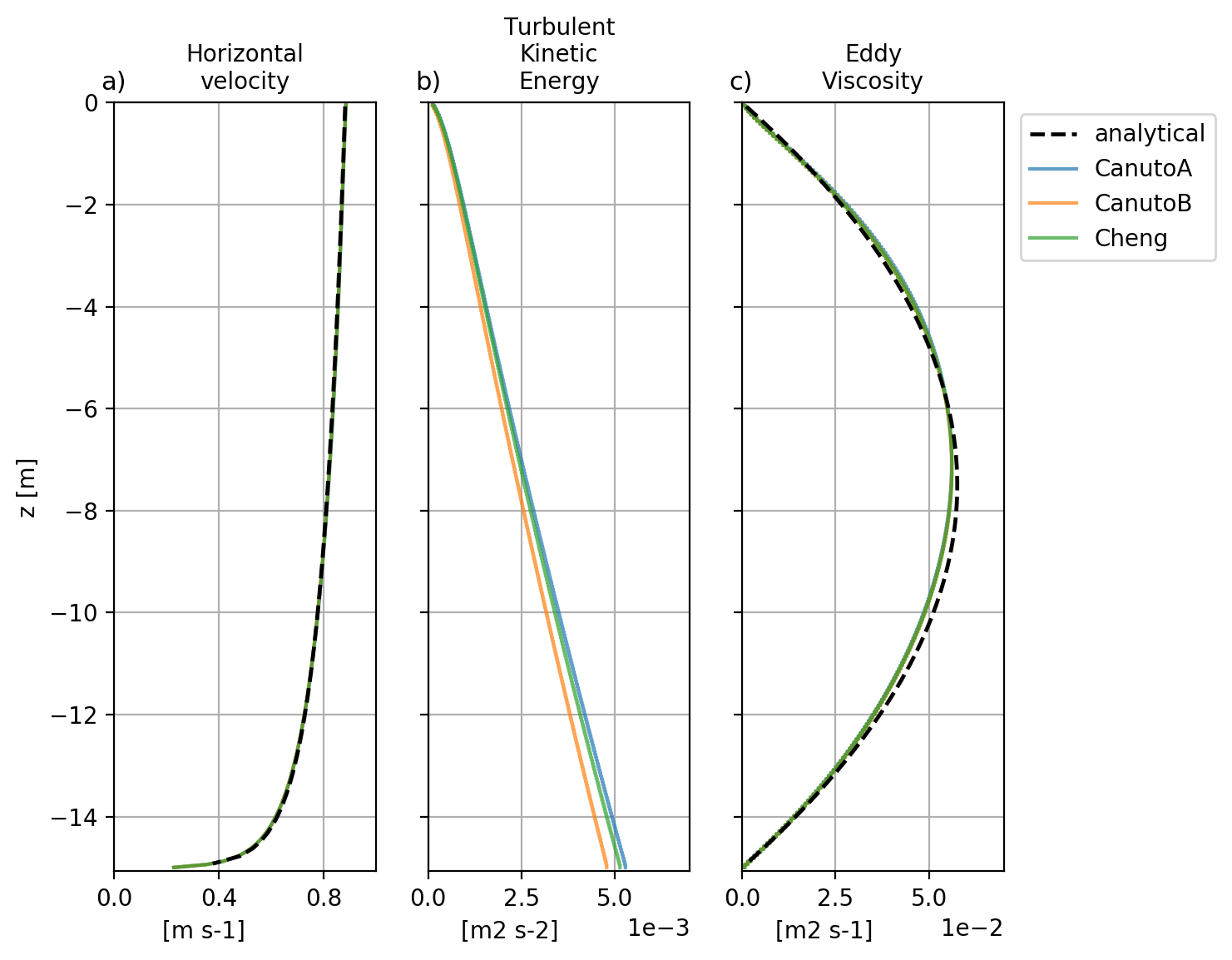}
  \caption
  {Bottom friction test with different stability functions.
  In all cases the $k-\omega$ closure were used.
  Canuto A and Cheng functions yield nearly identical results.
  Black dashed line stands for the analytical velocity and viscosity profiles.
  }\label{fig:bfriction_stab}
\end{figure}

Third, we experimented with the different stability functions.
We ran the $k-\omega$ model with Canuto A, B and Cheng stability
functions (Figure \ref{fig:bfriction_stab}).
In general the choice of the stability function does not have a major impact on
the results.
The results of Canuto A and Cheng stability functions are nearly identical.
Canuto B functions, on the other hand, yields slightly lower TKE, but the difference in viscosity and velocity is negligible.
The effect of stability functions was similar with other closure models as well
(not shown).

\subsection{Wind-driven entrainment} \label{sec:results_kato}

The next test examines mixed layer deepening due to surface stress, based on
the laboratory experiment originally conducted by \cite{kato1969}.
Initially the water column is at rest and linearly stratified.
A constant surface stress is applied at the free surface.
Stress induced mixing leads to a homogeneous surface layer that grows deeper
in time.
The depth of the mixed layer follows the empirical formula suggested by \cite{price1979}:

\begin{align}
 d_{\text{ML}} &= 1.05\ u^*_s \sqrt{\frac{t}{N_0}} \label{eq:kato_mldepth}
\end{align}
where $u^*_s$ is the surface friction velocity and $N_0$ is the initial
spatially invariable buoyancy frequency.
Here values $u^*_s  = 0.01\ \text{m}\ \text{s}^{-1}$ and
$N_0 = 0.01\ \text{s}^{-1}$ are used.
$N_0$ is prescribed by imposing a suitable linear salinity field and
using a linear equation of state.

The mixed layer deepening was simulated in 5 km wide square domain with
constant 50 m depth and 2.5 km mesh resolution.
All simulations were carried out with 30 s time step unless otherwise mentioned.

\begin{figure}[ht]
  \centering
  \noindent\includegraphics[width=0.6\textwidth]{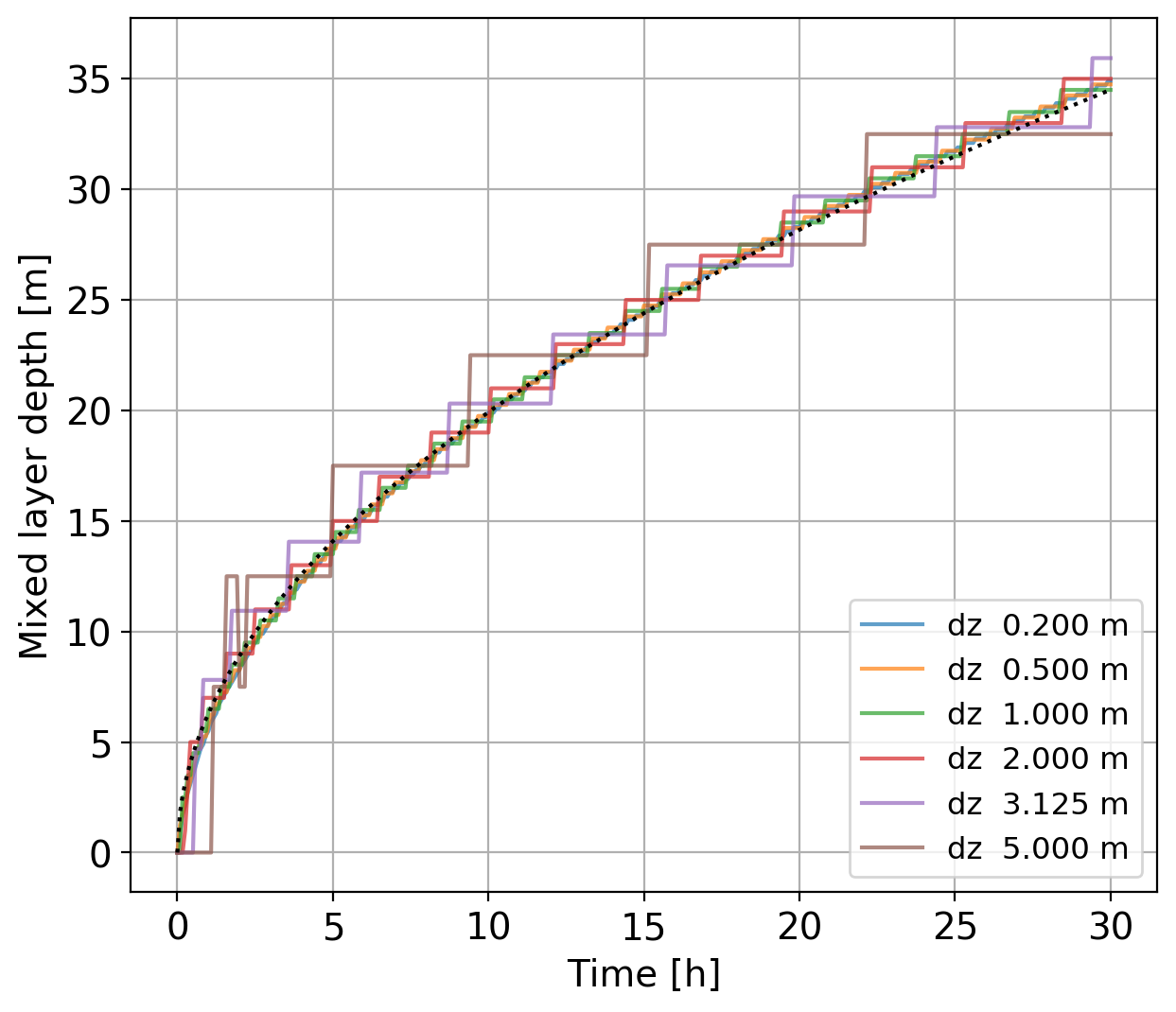}
  \caption
  {
  Mixed layer depth in wind-driven entrainment test with different mesh
  resolutions.
  The mixed layer depth was defined as the highest point where $k > 10^{-5}\ \text{m}^{2}\ \text{s}^{-2}$.
  In all cases the $k-\varepsilon$ model and Canuto A stability functions were used.
  Black dashed line stands for the empirical formula \eqref{eq:kato_mldepth}.
  }\label{fig:kato_resotest}
\end{figure}

\begin{figure}[ht]
  \centering
  \noindent\includegraphics[width=0.6\textwidth]{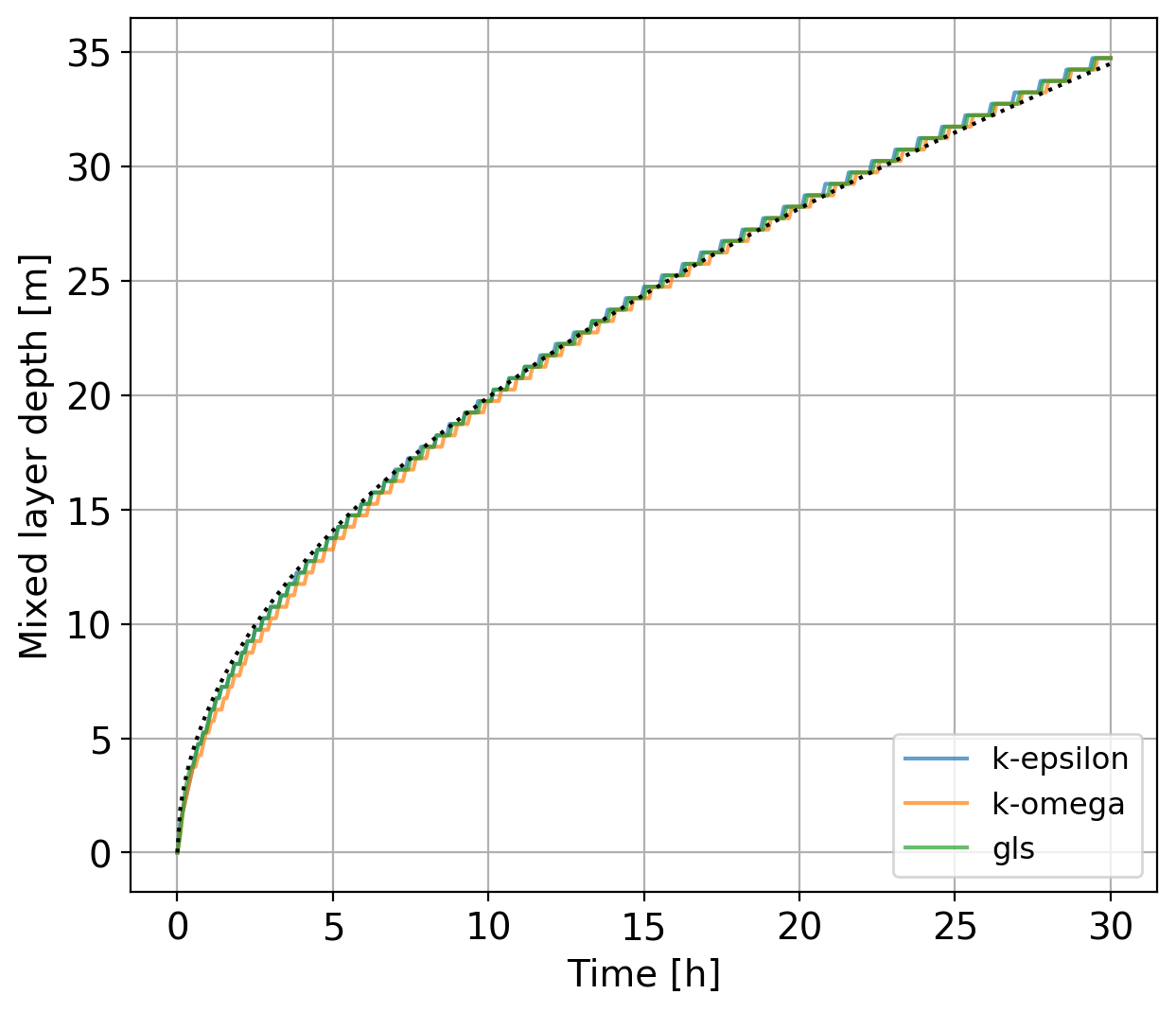}
  \caption
  {
  Mixed layer depth in wind-driven entrainment test with different turbulence
  closures.
  The blue, yellow, and green line stand for $k-\varepsilon$, $k-\omega$, and GLS models, respectively.
  In all cases the Canuto A stability functions were used.
  Black dashed line stands for the empirical formula \eqref{eq:kato_mldepth}.
  }\label{fig:kato_glstest}
\end{figure}

Six different mesh resolutions, $\Delta z=0.2, 0.5, 1.0, 2.0, 3.125$, and $5.0\ \text{m}$,
were experimented with, resulting in 250, 100, 50, 25, 16, and 10 vertical levels,
respectively.
The corresponding parabolic Courant number ranges from 7 to 0.03.

The evolution of the mixed layer depth is presented in
Figure \ref{fig:kato_resotest} for $k-\varepsilon$ model and Canuto A stability
functions.
All resolutions result in mixed layer deepening that is in good agreement
with the empirical formula \eqref{eq:kato_mldepth}.
With low vertical resolution, the mixed layer depth advances in steps as mixing
penetrates new elements.

Next we tested the three different turbulence closure models
(with Canuto A stability functions) on the medium resolution $\Delta z=0.5\ \text{m}$ mesh (Figure \ref{fig:kato_glstest}).
All models reproduce a realistic mixed layer depth.
The $k-\varepsilon$ and GLS models are close to the empirical mixed
layer depth.
The $k-\omega$ model tends to under estimate the mixed layer depth, especially
in the beginning of the simulation, but the difference is small.
These results are in good agreement with previous studies (e.g., \citealt{warner2005,karna2012}).

We also carried out experiments with different stability functions (not shown);
in all cases, the mixed layer depth behaved correctly, its variance being
similar to what is seen in Figure \ref{fig:kato_glstest}.

\begin{figure}[ht]
  \centering
  \noindent\includegraphics[width=0.6\textwidth]{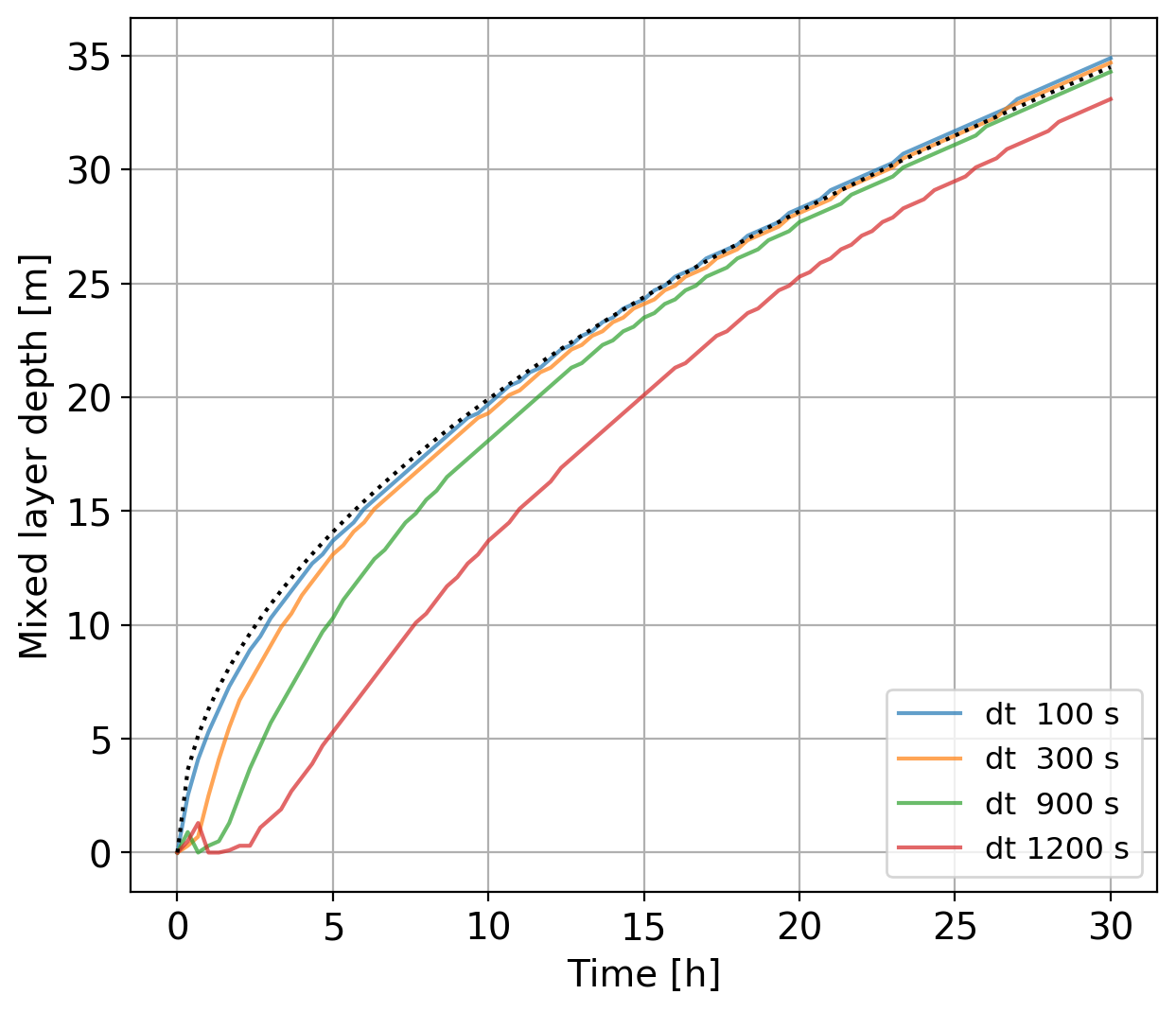}
  \caption
  {
  Mixed layer depth in wind-driven entrainment test with different time step.
  In all cases, the $k-\varepsilon$ model and Canuto A stability functions were used.
  Black dashed line stands for the empirical formula \eqref{eq:kato_mldepth}.
  }\label{fig:kato_dttest}
\end{figure}

The above tests were carried out with a short time step (30 s) close to typical values in regional applications.
To test how the model can handle stiff problems we increased the time step in the fine grid ($\Delta z=0.2\ \text{m}$) case up to $\Delta t = 1200\ \text{s}$ resulting in parabolic Courant number of 300 (Figure \ref{fig:kato_dttest}).
In all the presented cases, the model remains stable and produces a deepening mixed layer.
With long time steps ($>300\ \text{s}$), however, the deepening of the mixed layer is underestimated especially during the first 10 h of the simulation.
These results suggests that the scheme is indeed stable for stiff problems but starts to dissipate rapid processes (which occur mainly on the onset of the mixed layer in the beginning).
The observed behavior is consistent with results by \cite{reffray2015} where the NEMO 1D model also began to underestimate the mixed layer depth for very long time steps.

\subsection{Idealised estuary} \label{sec:results_estuary}

\begin{figure}[ht!]
  \centering
  \noindent\includegraphics[width=0.8\textwidth]{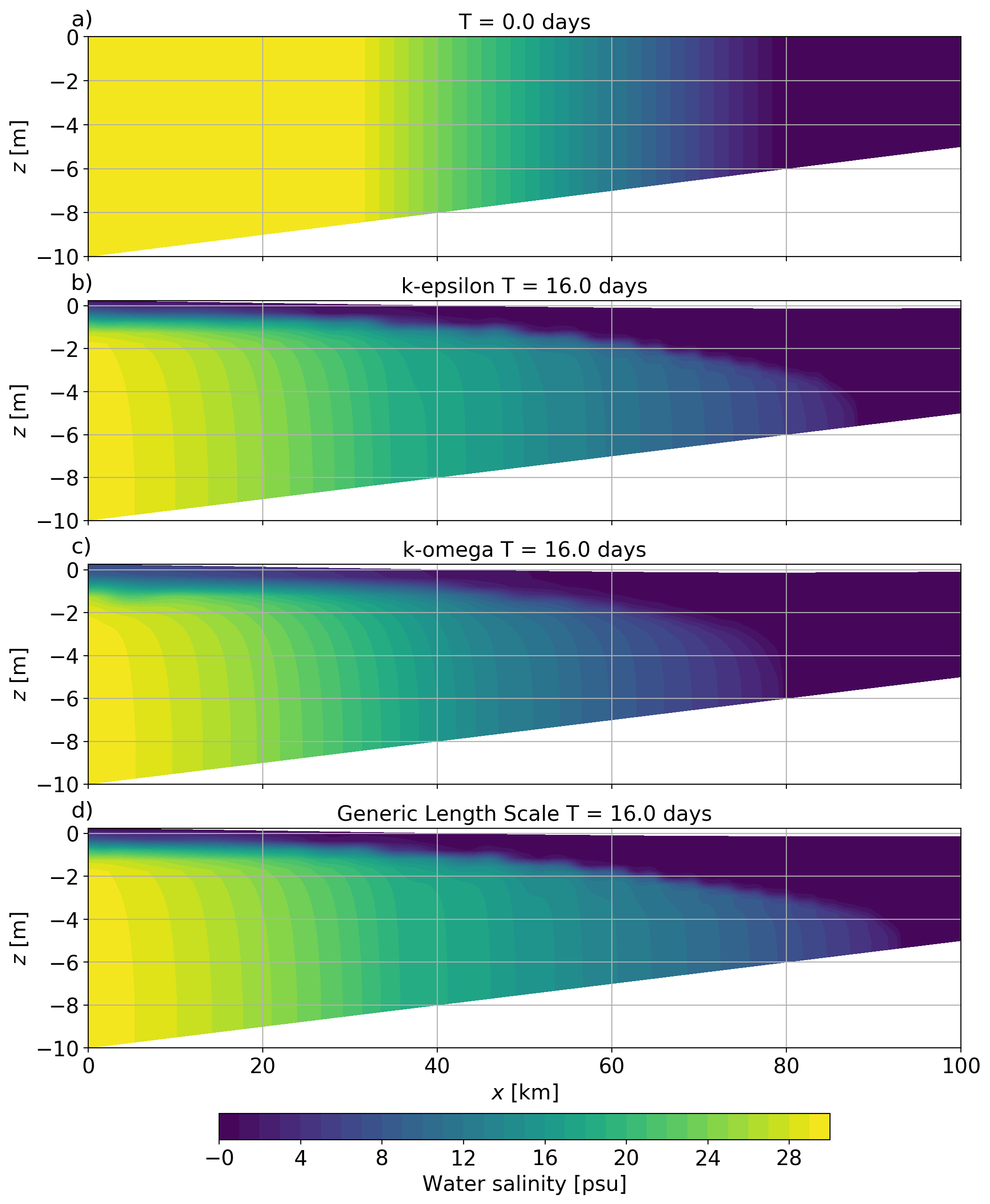}
  \caption
  {
  Idealized estuary simulation.
  a) Initial condition for salinity.
  b) Salinity field after 16 days of simulation for the $k-\varepsilon$ (b), $k-\omega$ (c), and GLS model (d).
  Canuto A stability functions were used in all simulations.
  }\label{fig:warner_fields}
\end{figure}

Estuarine circulation is an essential feature in coastal domains.
It is dominated by the interplay of buoyancy driven stratification and vertical
mixing.
Buoyancy input from a river tends to tilt isopycnals and increase
stratification in the estuary.
On the other hand, bottom friction and up-estuary currents tilt the isopycnals in the opposite direction causing unstable stratification in the bottom layer.
This generates vigorous mixing at the bed which, under suitable conditions, can mix the entire water column.
Under sufficiently strong stratification, however, turbulence cannot penetrate the
pycnocline resulting in a well mixed bottom layer that is de-coupled from
the fresher surface layer.
The competition between these two tendencies, and temporal evolution of the turbulence fields, cannot be modeled without
a dynamic two-equation turbulence closure model.

\begin{figure}[ht!]
  \centering
  \noindent\includegraphics[width=0.9\textwidth]{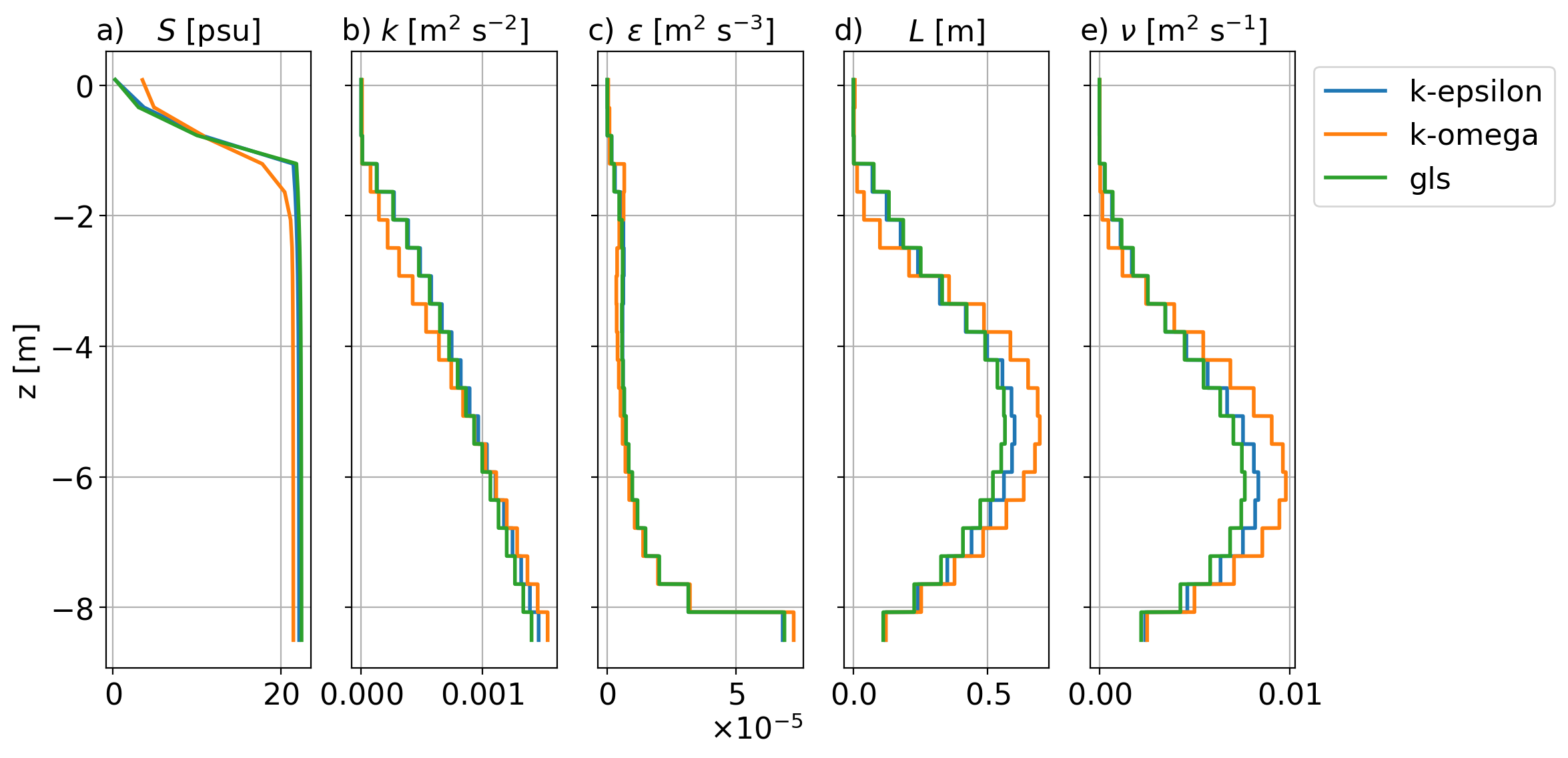}
  \caption
  {
  Idealized estuary simulation.
  Vertical profiles after 14.40 days of simulation at $x = 30\ \text{km}$ from
  the mouth.
  Canuto A stability functions were used in all simulations.
  }\label{fig:warner_profiles}
\end{figure}

We test the model's capability of representing estuarine circulation
with the idealized estuary test case by \cite{warner2005}.
This test case has been used to validate other models as well, e.g.,
SLIM \citep{karna2012}, and SELFE \citep{lopez2017}.

The idealized estuary test case does not have an analytical solution.
Furthermore, our tests suggest that the numerical solution is highly sensitive to the open boundary conditions which vary in each model due to its discretization.
As such, carrying out a direct comparison of the modeled fields is not beneficial.
Nevertheless, this is a useful dynamical test to verify that the coupled model can generate realistic estuarine circulation features;
It also reveals possible numerical instabilities associated with strong horizontal gradients, advection, and tidal currents.

The domain is a rectangular channel 100 km long and 1 km wide. Depth varies
linearly from 10 m at the deep (ocean) boundary to 5 m in the shallow (river)
end.
Horizontal mesh resolution is 500 m; 20 equally distributed vertical levels are used.

At the river boundary, a constant discharge of
$F_{river} = 400\ \text{m}^3\ \text{s}^{-1}$
is imposed, corresponding to a 8 $\text{cm}\ \text{s}^{-1}$ seaward velocity.
Water elevation is unprescribed, and water salinity is set to zero.
At the ocean boundary, a sinusoidal tidal current is imposed with
$0.4\ \text{m}\ \text{s}^{-1}$ amplitude and period $T_{tide}=12\ \text{h}$.
In addition, we superimpose the (seaward) river flux, $F_{river}$, at ocean boundary
as well to ensure tidally averaged volume conservation.
Salinity of in-flowing ocean water is set to is 30 psu.
Throughout the simulation temperature is kept at constant $10\ ^\circ\text{C}$.

Initially water is at rest.
Salinity varies from 30 to 0 psu along the channel between 30 and 80 km from
the ocean boundary (Figure \ref{fig:warner_fields} a).
The ocean and river boundary conditions are ramped up for one hour.
During the simulation estuarine circulation quickly develops, driving
freshwater seaward in the surface layer.
A thick salt wedge forms, oscillating with the tides.
The model is spun up for 15 days after which the solution has reached a
quasi-periodic state where each tidal cycle is nearly identical.

Figure \ref{fig:warner_fields}, shows the instantaneous salinity
distribution after 16 days of simulation for $k-\varepsilon$ (b),
$k-\omega$ (c) and GLS models (d). In all simulations the Canuto A stability
functions were used.
The results show a thick salt wedge, and a fairly thin, seaward flowing surface
layer.
Salinity reaches about 80 km from the estuary mouth.
The $k-\omega$ model shows lower stratification and shorter salinity intrusion
indicating stronger effective mixing.
The GLS model results in the largest salinity intrusion.

Vertical profiles are presented in Figure \ref{fig:warner_profiles} for the
three closures.
The profiles indicate that the well-mixed bottom layer fills most of the water
column.
In the upper water column stratification is very strong which limits mixing and
results in almost turbulence-free surface layer.
Salinity profiles in Figure \ref{fig:warner_profiles} (a) show that
stratification is indeed weaker in the $k-\omega$ model.
It also generates higher TKE and viscosity in the bottom layer suggesting a
larger overall (tidally-averaged) mixing.
The $k-\varepsilon$ and GLS model behave quite similarly.
GLS, however, generates lower TKE and viscosity which is consistent with the
larger salinity intrusion length (Figure \ref{fig:warner_fields}).

Qualitatively the presented results are similar to those in
\cite{warner2005}, \cite{karna2012} and \cite{lopez2017}.
Some differences do exist in salt intrusion and stratification, for example,
which could be attributed to differences in the ocean boundary
conditions and, more generally, the model's discretization.
It should also be noted that the behavior of the three turbulence closure
models strongly depends on the chosen parameters.

\subsection{Application to Columbia River plume} \label{sec:results_creplume}

\begin{figure}[ht]
  \centering
  \noindent\includegraphics[width=1.0\textwidth]{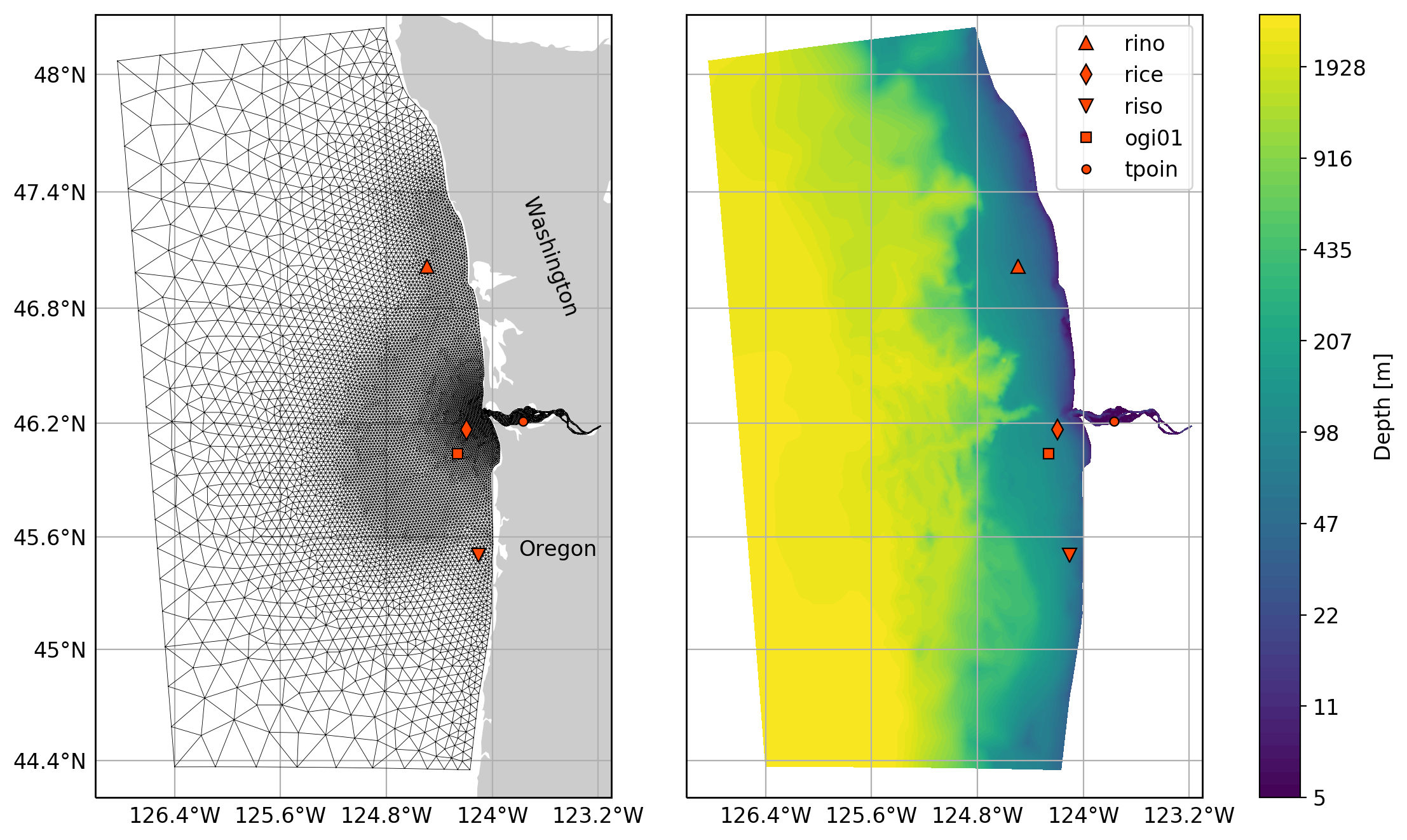}
  \caption
  {
  Columbia River plume model domain.
  Horizontal mesh (left) consist of 17 000 triangles whose size ranges between
  900 m and 15 km.
  The bathymetry (right) reaches 2940 m in the deepest part of the domain.
  In the vertical direction 20 terrain-following vertical levels are used,
  resulting in 340 000 prisms and $2\times10^6$ tracer degrees of freedom.
  Vertical resolution is concentrated near the surface;
  the first layer is roughly 1 m thick in the plume region.
  }\label{fig:cre_mesh}
\end{figure}

The Columbia River is a major freshwater source in the west coast of the
United States (Figure \ref{fig:cre_mesh}).
With an annual mean discharge of 5500 $\text{m}^3\ \text{s}^{-1}$
it is the second largest river in continental USA \citep{chawla2008}.
Maximal daily tidal range varies between from less than 2.0 to 3.6 m at the
river mouth \citep{karna2015}.
The river plume is a major feature in the Pacific Northwest Coast,
affecting stratification, currents, and nutrients in the coastal
waters \citep{hickey2003}.

The river plume can be divided into different water masses.
\cite{hornerdevine2009} define four water masses:
the source water, followed by the tidal, re-circulating, and far-field plume.
The tidal plume (with 6 to 12 h time scale) consists of a freshwater lens being
emitted from the river mouth every tidal cycle.
Once released, it rapidly spreads out and becomes thinner, its depth ranging
from 6 to 3 m.
Salinity is below 21 psu.
The tidal plume eventually merges with the re-circulating plume.

The re-circulating plume forms an anticyclonic bulge at the vicinity of the
river mouth.
In the cross-shore direction, the bulge is roughly 40 km in diameter.
The re-circulating plume consists of freshwater volume equivalent to 3–4 days of river discharge; salinity is between 21 and 24 psu.

The far-field plume lies beyond the re-circulating plume.
It is the zone where final mixing of the plume and ambient ocean waters takes
place, unaffected by the momentum of the river discharge.
Typically, the far-field plume forms a large coastal current,
driven by buoyancy and Earth's rotation, that can extend hundreds of
kilometers North from the river mouth.
The evolution of the far-field plume, however, is strongly affected by
winds and coastal currents.

The typical northward propagation of the river plume can be arrested by
northerly, upwelling-favorable winds.
Upwelling generates an off-shore current in the surface layer, eroding the
river plume from the coast.
The plume detaches from the coast, spreads out and becomes thinner.
Under sufficiently strong and prevailing winds, the plume begins to travel
southwest.
Northerly winds are more common in the Pacific Northwest during summer months,
which is why the southwest-oriented plume is often called the summer plume.
Upwelling-favorable winds, therefore, can reverse the orientation of the
plume.

Southerly, downwelling-favorable winds, on the other hand, enhance the
northward coastal current.
Under downwelling conditions, the surface layer flows onshore and pushes the
plume against the coast,
making it narrower and thicker.
Consequently, the coastal current becomes more pronounced:
it becomes narrower and propagates faster towards the North.

The evolution of the river plume was simulated with Thetis.
The computational domain spans from roughly 40 $^\circ$N to 50 $^\circ$N in the
zonal direction
(Figure \ref{fig:cre_mesh}).
The western boundary is located roughly 100 km west of the coast.
The Columbia River is included in the domain, excluding the shallow
lateral bays.
The river boundary is located at Beaver Army, 86 km upstream from the
mouth.
The horizontal mesh resolution (triangle edge length) ranges from roughly 900 m
in the estuary to 15 km at the open ocean boundaries.
In the vertical direction, 20 terrain-following layers were used.
The layer thickness ranges from roughly 1 m near the surface to a maximum of
180 m at the bottom.
The entire vertical grid adjusts to the free surface similarly to $z^*$
grid used in structured grid models \citep{adcroft2004}.

We used a composite bathymetry data \citep{karna2016b} that was interpolated on
the model grid (Figure \ref{fig:cre_mesh}).
In addition, the bathymetry was smoothed with a Laplacian diffusion operator to
limit strong gradients which could introduce internal pressure errors.
Minimum water depth was set to 3.5 m.
Wetting and drying was neglected.

Atmospheric pressure and 10 m wind were obtained from the North American Mesoscale Model (NAM) model.
Wind stress was computed with the formulation by \cite{large1981}.
At the ocean boundary temperature, salinity and sub-tidal velocity was imposed
from the global Navy Coastal Ocean Model (NCOM) model \citep{barron2006}.
In addition, tidal elevation and velocities were imposed from the TPXO v9.1 model \citep{egbert2002}.
At the river boundary discharge and water temperature are imposed using
US Geological Survey (USGS) gauge data. Salinity is set to zero.

The simulation covers a time period from May 1 2006 to July 2.
The first month is used to spin up the simulation;
the analysis period is from June 1 onward.
This spin-up time is sufficient for the estuary–plume system as
the estuary has a residence time of a few days \citep{karna2016a}
and the far-field plume develops in a couple of weeks.

It is worth noting that the Thetis model setup has not been fully calibrated for the system and the results are therefore preliminary.
An exhaustive model calibration and skill analysis are out of the scope of the present article and will be addressed in the future.

The simulations were carried out on a Linux cluster with 16-core
Intel Xeon E5620 processors and Mellanox Infiniband interconnect.
96 MPI processes were used.
The simulation had $2\times10^6$ tracer degrees of freedom (DOF) and
used 15 s time step; the full 90-day simulation took 7.2 days.

\begin{figure}[ht]
  \centering
  \noindent\includegraphics[width=0.9\textwidth]{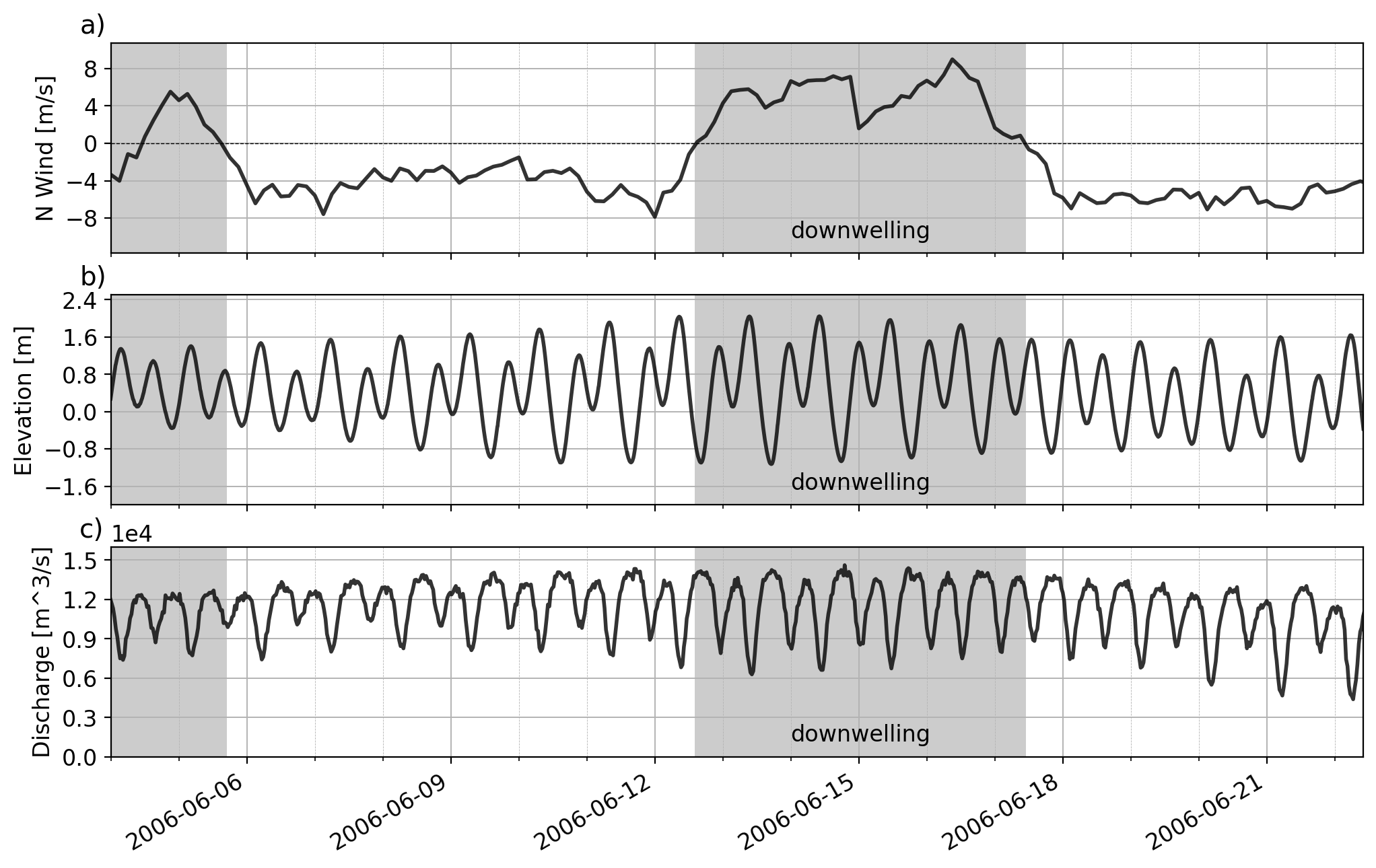}
  \caption
  {
  Time series of wind, tidal and river forcings in the Columbia River plume
  application.
  The northward wind component (a) is from the atmospheric model at
  the rice station;
  The tidal elevation (b) is from tpoin tide gauge;
  River discharge (c) is from Beaver Army gauge.
  The shaded areas indicate strong downwelling periods.
  }\label{fig:cre_forcings}
\end{figure}

Figure \ref{fig:cre_forcings} shows the atmospheric and fluvial forcing
conditions for the analysis period.
Winds alternate between up- (northerly) and downwelling-favorable direction
(Figure \ref{fig:cre_forcings} a).
Tides exhibit a typical spring-neap variation to the system, tidal range varying between 1.6 and 3.4 m;
River discharge is roughly 10000 $\text{m}^3\ \text{s}^{-1}$ during the simulation (Figure \ref{fig:cre_forcings} b and c).

\begin{figure}[ht]
  \centering
  \noindent\includegraphics[width=0.9\textwidth]{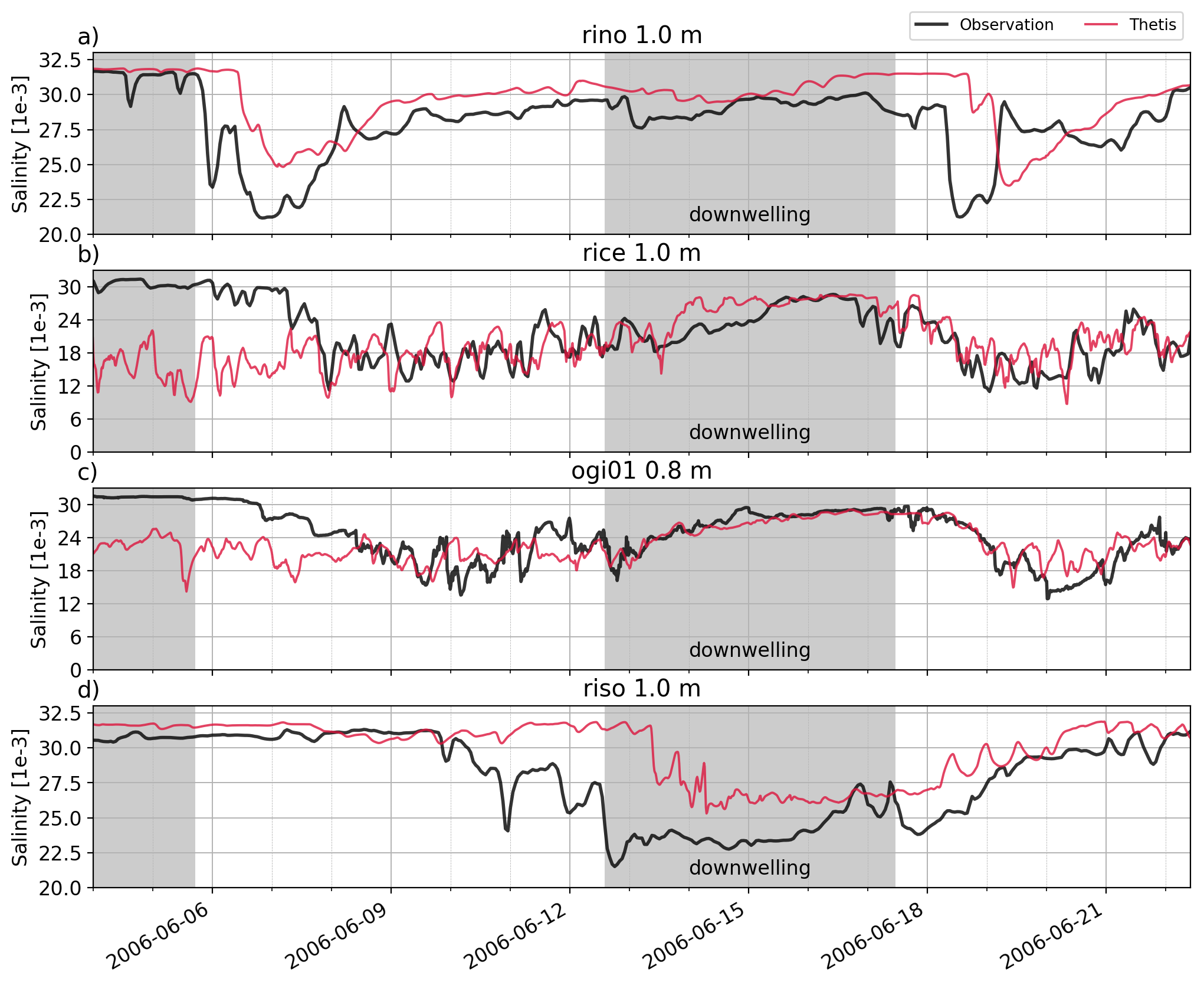}
  \caption
  {
  Time series of observed and modeled surface salinity at selected stations in the
  Columbia River application.
  The shaded areas indicate strong upwelling periods.
  The station locations are shown in Figure \ref{fig:cre_mesh}.
  }\label{fig:cre_station_salt}
\end{figure}

Simulated salinity field is compared against an observational data set from the RISE project (River Influences on Shelf
Ecosystems, e.g., \citealt{maccready2009}).
\cite{maccready2009} and \cite{liu2009} use data from the 2004 summer measurement campaign.
It should be noted that for the 2006 campaign, used in the present analysis, the north and south moorings were moved further out from the river mouth by some 80 km (see Figure \ref{fig:cre_mesh} for the mooring locations).
Therefore, the 2006 data set is well-suited for evaluating the behavior of the coastal current and the far-field plume under changing wind conditions.

Figure \ref{fig:cre_station_salt} shows time series of observed surface
salinity at stations rino, rice, ogi01, and riso.
The rice  station (Figure \ref{fig:cre_station_salt} b) is located in the tidal plume recording the passing of each
tidal freshwater lens.
The model replicates sub-tidal and tidal evolution variability of the near field plume well.
Note that exact replication of the tidal signal is difficult as salinity gradients are very large.
Measured salinity therefore strongly depends on the passing of the plume fronts, which is sensitive to atmospheric forcing, mesh resolution, mixing, and bathymetry features, for instance.

The ogi01 station (Figure \ref{fig:cre_station_salt} c) is located some 25 km southeast from the river mouth.
It captures the re-circulating plume when it is pushed southward from the river mouth.
Typically this occurs during southerly winds.
During the simulation period, two of such events are recorded, on June 6--10 and June 18--20.
The model captures the southward transport of the plume relatively well, especially during the latter event.

The northern station, rino, is located in the far-field plume region, 24 km off the coast (Figure \ref{fig:cre_station_salt} a).
The narrow coastal current, however, is often situated closer to the coast.
The plume passes through rino station only in cases where the winds erode the plume from the coast.
Such events are seen in the observations on June 6 and 18.
The model captures those events well, although the arrival of the plume is delayed by roughly one day.
Salinity in Thetis also tends to be higher compared to the observations.

The southern station, riso, captures the wind-driven southward traversal of the plume. The observations show such an event between June 10 and 21.
Thetis simulation shows the passage of the plume, albeit the onset is delayed by roughly 3 days, and surface salinity does not drop as much as in the observations.

\begin{figure}[ht]
  \centering
  \noindent\includegraphics[width=0.9\textwidth]{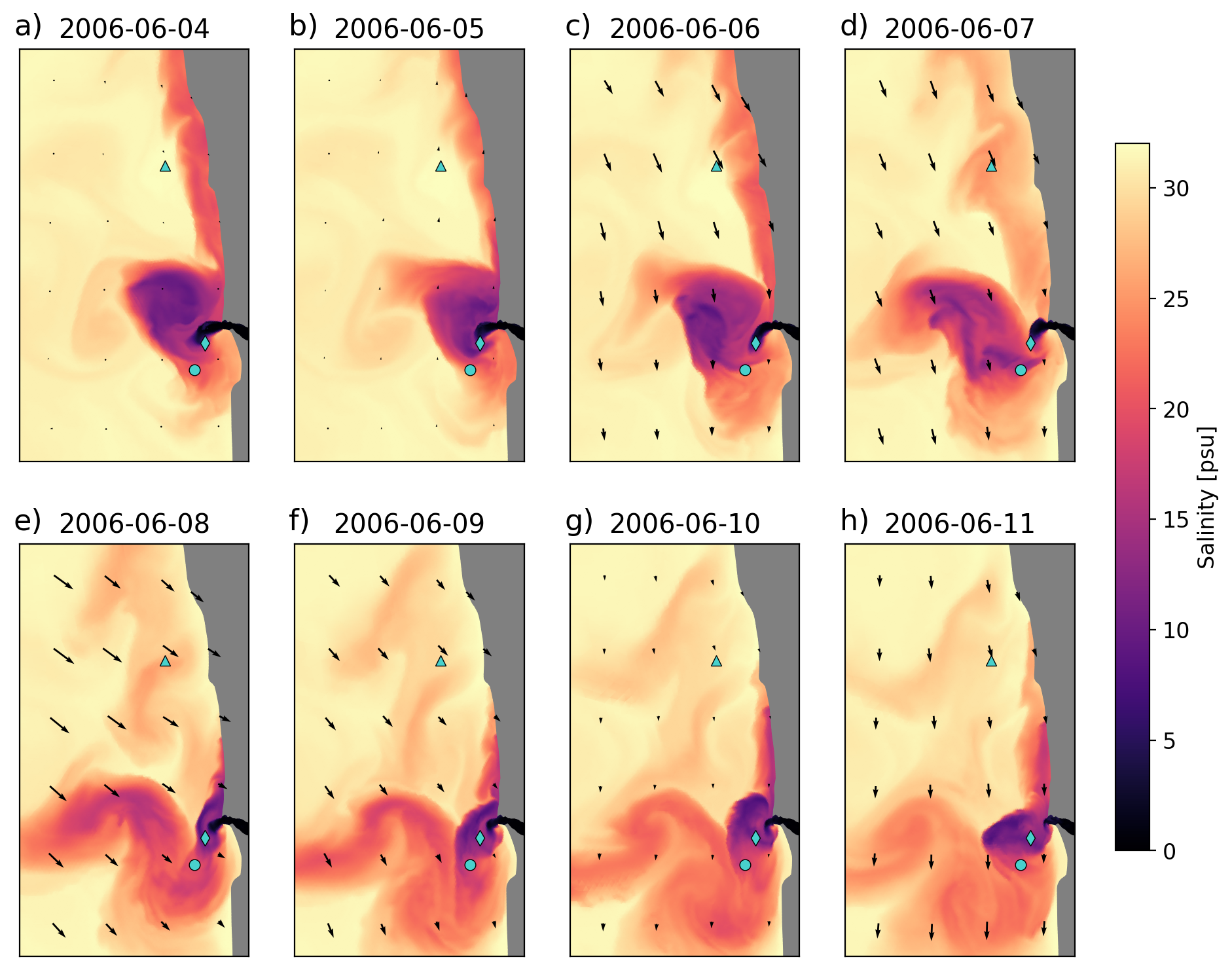}
  \caption
  {
  Evolution of surface salinity in the Columbia River application.
  The plotted stations are rino ($\blacktriangle$), rice ($\blacklozenge$),
  ogi01 ({\large$\bullet$}).
  }\label{fig:cre_sss_evolution}
\end{figure}

Figure \ref{fig:cre_sss_evolution} shows the evolution of the surface salinity
in the Thetis simulation.
Prior to the first event (panels a and b),
the bulge of the re-circulating plume is clearly visible just north of the river mouth.
The far-field plume is quite narrow and fresh; it travels northward by the coast and does not reach the rino station.
On the onset of northerly winds (panels c and d), the far-field plume detaches from the coast, spreads out, and passes through the station.
As the winds prevail, the plume spreads out more and starts to travel southwest (panels e - h).
Eventually, a new coastal plume starts to develop.
Note that in response to the winds, the re-circulating plume also spreads out, dilutes and begins to meander.

\section{Discussion} \label{sec:discussion}

The idealized 1D and estuary test cases show that the GLS implementation remains stable and produces realistic bottom and surface layer dynamics as well as estuarine circulation.
Overall the model's performance is in good agreement with results presented in
the literature (e.g., \citealt{warner2005,karna2012}).

All the tested closure models and
stability functions yield similar results, although some differences do appear.
For example, the differences in the bottom friction test case between different
closures (Figure \ref{fig:bfriction_closure}) are mostly due
to numerics related to the surface boundary conditions.
The commonly-used choice of $k-\varepsilon$ and Canuto A stability functions,
however, appears to perform well in all test cases.
The model performance is also strongly dependent on the chosen parameters
which can be tuned in applications.
For example, vertical profiles in the bottom friction test are affected by
the choice of $c_{\mu}^0$ and $\sigma_\Psi$ parameters, among others.
Similarly, the length of salinity intrusion in the idealized estuary test case
(Figure \ref{fig:warner_fields}) depends on the magnitude of total mixing which
is partially controlled by the minimum value of TKE, $k_{min}$.
Evidently, the parameters could be tuned further to reduce the gap between
different closure models, for example.
However, such an investigation of the parameter space is out of the scope of
the present study.

The Columbia River plume simulation indicates that Thetis can represent tidal,
highly dynamic river plumes.
The plume features numerous eddies and responds quickly to changing tidal and wind conditions.
Overall the simulation is in good agreement with the observations and known behavior of the Columbia plume under up- and downwelling conditions \citep[e.g.,][]{hickey2003}.
Qualitatively, Thetis results compare well with ROMS and SLIM \citep{vallaeys2018}.
Thetis surface fields show eddies and filaments, and strong frontal features are retained over several tidal cycles.
This suggests that numerical mixing is low, as \cite{karna2018a} demonstrated with idealised test cases.

The presented model uses $\text{P}^{\text{DG}}_0$ function space for the
turbulent quantities while the mean flow values are in $\text{P}^{\text{DG}}_1$
space.
This choice resembles the staggered discretization used in finite volume models
(e.g., in GOTM) where the mean flow and turbulent variables are defined at cell
centers and cell interfaces, respectively.
As the turbulent quantities reside in a lower-degree function space,
it could impact the accuracy of the coupled model.
However, as the turbulence closure model only affects the mean flow model
through the eddy viscosity and diffusivity fields, the impact is likely to be
small.
Deriving a higher-order discretization for the turbulent quantities remains a
topic for future research.
Similarly, ensuring strict energy conservation between the turbulent and mean flow fields will be addressed in the future.

\section{Conclusions} \label{sec:conclusions}

We have presented a DG finite element discretization for the GLS turbulence
closure model and a positivity-preserving coupled time integration scheme.
The turbulence closure model was implemented in the Thetis circulation model.

The model was tested with a series of idealized test cases.
The results verify that bottom boundary layer, surface mixing, and
estuarine circulation processes are simulated correctly.
All the three closures ($k-\varepsilon$, $k-\omega$, and GLS) and stability
functions (Canuto A, B, and Cheng) yield similar performance.

Finally, the model was tested in a realistic Columbia River plume
application.
The results indicate that the coupled model performs
well in complex, highly dynamic plume applications.
The surface salinity values are in good agreement with observations and the plume responds to changing wind
forcing.
The plume shows strong frontal features
which suggests low numerical diffusion.
Thetis, therefore, can retain fronts between water masses and capture the
observed plume traversals.

There is an ever-increasing need for accurate, less dissipative coastal ocean
models that can simulate buoyancy-driven coastal flows with a realistic
representation of frontal features and vertical mixing.
As such, the presented coupled circulation-turbulence model constitutes an
important step towards next-generation ocean modeling.

\section*{Acknowledgments}

I am grateful to
anonymous reviewers whose insightful comments helped to improve the manuscript.
Lawrence Mitchell (Durham University),
Stephan Kramer and David Ham (Imperial College London) are thanked
for their contributions to Thetis and continuous support with Firedrake.
Antonio Baptista (Oregon Health \& Science University) is acknowledged for his
contribution on the Columbia River plume application.


\appendix

\section{Stability functions}\label{sec:stability-functions}

All the non-equilibrium stability functions are defined by a set of canonical parameters \citep[Table 1 in][]{umlauf2005}.
The functions can, however, be expressed as rational functions of $\alpha_M$ and $\alpha_N$ with parameters $n_i, d_i, n_{bi}$ \citep{umlauf2005},

\begin{align}
  c_\mu &= \frac{n_0 + n_1 \alpha_N + n_2 \alpha_M}{d_0 + d_1 \alpha_N + d_2 \alpha_M + d_3 \alpha_N \alpha_M + d_4 \alpha^2_N + d_5 \alpha^2_M}. \label{eq:stability_visc}\\
  c'_\mu &= \frac{n_{b0} + n_{b1} \alpha_N + n_{b2} \alpha_M}{d_0 + d_1 \alpha_N + d_2 \alpha_M + d_3 \alpha_N \alpha_M + d_4 \alpha^2_N + d_5 \alpha^2_M}. \label{eq:stability_diff}
\end{align}

\cite{canuto2001} introduced two sets of full non-equilibrium stability functions, usually referred to as Canuto A and B.
The stability region of Canuto B functions is somewhat larger than that of version A, suggesting that it is more robust in practice \citep{umlauf2005}.
Another stability function, based on a very similar formulation, is given by \cite{cheng2002}.
All the three stability function families have been implemented using the formulation \eqref{eq:stability_visc}-\eqref{eq:stability_diff}.

\subsection{Limiting $\alpha_M$ and $\alpha_N$}\label{sec:limit-stability-funcs}

The stability functions are only defined for a certain values of $\alpha_N$ and $\alpha_M$ due to numerical and physical reasons.
First, the shear frequency $\alpha_M$ is positive by definition. To ensure its positivity $\alpha_N$ must be limited from below.
The condition $\alpha_M=0$ corresponds to the equilibrium relation $B = \varepsilon$, for which it holds \citep[eq. A.23 in][]{umlauf2005}
\begin{align}
  \frac{B}{\varepsilon} = - c'_\mu(\alpha_N) \alpha_N = 1. \label{eq:eq_pos_an}
\end{align}
This equilibrium condition corresponds to unstably stratified, convective regime where $\alpha_N < 0$ and $B > 0$.
Substituting the stability function $c_\mu'$ into \eqref{eq:eq_pos_an} results in
\begin{align}
  \alpha_N^{\text{min}} = \frac{-(d_1 + n_{b0}) + \sqrt{(d_1 + n_{b0})^2 - 4 d_0 (d_4 + n_{b1})}} {2(d_4 + n_{b1})} \label{eq:alpha_n_min}.
\end{align}
Typical value of $\alpha_N^{\text{min}}$ is around -3.

An alternative constrain to $\alpha_N$ is to require monotonicity \citep[eq. 47 in][]{umlauf2005}:
\begin{align}
    \pd{}{\alpha_N}\left( \frac{c'_\mu(\alpha_N)}{\alpha_N} \right) > 0,
\end{align}
which yields more strict limit.
In this work \eqref{eq:alpha_n_min} is used as it was sufficient in all the tested cases.

Cropping $\alpha_N$ at the minimum value may lead to numerical
oscillations as the gradient of the stability function is large for
negative $\alpha_N$. Therefore a smoother transition has been proposed
\citep[eq. (19) in ][]{burchard1999}:
\begin{align}
  \tilde{\alpha}_N = \alpha_N - \frac{(\alpha_N - \alpha_c)^2}{\alpha_N + \alpha_N^{\text{min}} -2\alpha_c},\quad \forall \alpha_N < \alpha_c \label{eq:alpha_n_min_smooth}
\end{align}
The smoothness of the transition is controlled by $\alpha_c$, defined in range $\alpha_N^{\text{min}} < \alpha_c < 0$. Here a value $\alpha_c = -1.2$ is used.

In addition, the shear frequency, $\alpha_M$, must be limited above to ensure physically sound behavior and numerical stability \citep[eq. 44 in][]{umlauf2005,burchard2001a}:
\begin{align}
  \alpha_M^{\text{max}} \approx \frac{d_0 n_0 + (d_0 n_1 + d_1 n_0) \alpha_N + (d_1 n_1 + d_4 n_0) \alpha_N^2 + d_4 n_1 \alpha_N^3}
  {d_2 n_0 + (d_2 n_1 + d_3 n_0) \alpha_N + d_3 n_1 \alpha_N^2} \label{eq:alpha_m_max}
\end{align}
Note that $\alpha_M^{\text{max}}$ depends on $\alpha_N$.

The limit \eqref{eq:alpha_m_max} is not always sufficient to ensure stability of the numerical scheme;
\cite{deleersnijder2008} present a more generic stability constraint which guarantees that the vertical gradient of the turbulent quantities remains bounded.
In this work, \eqref{eq:alpha_m_max} is used as it has been sufficient for stability:
We first set $\alpha_N$ to $\tilde{\alpha}_N$ according to \eqref{eq:alpha_n_min_smooth}, and then apply the $\alpha_M^{\text{max}}$ limit.

\section{Choosing empirical parameters}\label{sec:choosing_parameters}

\subsection{Computing $c_3^-$}\label{sec:choosing_c3minus}

The parameter $c_3^-$ controls the buoyancy production of $\Psi$ under
stable stratification.
Its value can be computed from \eqref{eq:ri_steady_state}:
\begin{align}
  c_3^- = c_2 - (c_2 - c_1)\frac{c_\mu(R_i^{st})}{c'_\mu(R_i^{st})}\frac{1}{R_i^{st}} \label{eq:c3_minus}
\end{align}

The value of $c_3^-$ thus depends on both the closure (parameters $c_i$) and the stability functions.
To evaluate \eqref{eq:c3_minus} one first needs to evaluate the stability functions under steady state.
The steady state condition \eqref{eq:steady_k_source} can also be written as \citep[eq. (49) in][]{umlauf2005}

\begin{align}
  c_\mu \alpha_M - c'_\mu \alpha_N = 1, \label{eq:steady_k_source_alpha}
\end{align}

from which one obtains a non-linear equation
\begin{align}
  c_\mu(\alpha_M, R_i^{st}) \alpha_M - c'_\mu(\alpha_M, R_i^{st}) \alpha_M R_i^{st} &= 1. \label{eq:ri_st}
\end{align}
From the above equation $\alpha_M$ can be solved either analytically or numerically.
In practice, the analytical approach has been more robust and it is used in this work;
the values for the three closures and the Canuto A stability functions are shown in Table \ref{tab:gls_parameters}.

\subsection{Parameter $c_\mu^0$}\label{sec:cmu0}

Parameter $c_\mu^0$ stands for the value of the stability function in
unstratified equilibrium where $P=\varepsilon$.

From \eqref{eq:steady_k_source_alpha} and \eqref{eq:gradient_ri} it follows that in equilibrium the stability functions only depend on $R_i^{st}$:
$c_\mu = c_\mu(R_i^{st}), c'_\mu = c'_\mu(R_i^{st})$.
Its value can therefore be computed for each stability function
\citep[see eq. A.22 in ][]{umlauf2005}.

\subsection{Parameter $\sigma_\Psi$}\label{sec:sigmapsi}

A relation for the Schmidt number, $\sigma_\Psi$, and the von Karman constant
$\kappa$ can be derived for the logarithmic boundary layer assumption
\citep[eq. 14 in][]{umlauf2003b}:

\begin{align}
 \sigma_\Psi &= \frac{n^2 \kappa^2}{(c_\mu^0)^2 (c_2 - c_1)} \label{eq:sigma_psi}
\end{align}

In this work we compute $\sigma_\Psi$ for each closure with $\kappa=0.4$.

\section{Source code availability} \label{sec:source_code}

The source code used to perform the experiments in
this papers is publicly available.
Firedrake, and its components,
may be obtained from \texttt{https://www.firedrakeproject.org/}
(last access: 1 February 2020);
Thetis from \texttt{http://thetisproject.org/} (last access: 1 February 2020).
For reproducibility, we also cite archives of the exact software versions used
to produce the results in this paper.
All major Firedrake components and the Thetis source code have been archived on Zenodo \citep{zenodo/Firedrake,zenodo/Thetis}.


 \bibliographystyle{elsarticle-harv} 
 \bibliography{references}






\end{document}